\documentclass[twocolumn]{aa}
\usepackage[draft]{minted}

\makeatletter
\def\PYGdefault@reset{\let\PYGdefault@it=\relax \let\PYGdefault@bf=\relax%
    \let\PYGdefault@ul=\relax \let\PYGdefault@tc=\relax%
    \let\PYGdefault@bc=\relax \let\PYGdefault@ff=\relax}
\def\PYGdefault@tok#1{\csname PYGdefault@tok@#1\endcsname}
\def\PYGdefault@toks#1+{\ifx\relax#1\empty\else%
    \PYGdefault@tok{#1}\expandafter\PYGdefault@toks\fi}
\def\PYGdefault@do#1{\PYGdefault@bc{\PYGdefault@tc{\PYGdefault@ul{%
    \PYGdefault@it{\PYGdefault@bf{\PYGdefault@ff{#1}}}}}}}
\def\PYGdefault#1#2{\PYGdefault@reset\PYGdefault@toks#1+\relax+\PYGdefault@do{#2}}

\expandafter\def\csname PYGdefault@tok@w\endcsname{\def\PYGdefault@tc##1{\textcolor[rgb]{0.73,0.73,0.73}{##1}}}
\expandafter\def\csname PYGdefault@tok@c\endcsname{\let\PYGdefault@it=\textit\def\PYGdefault@tc##1{\textcolor[rgb]{0.25,0.50,0.50}{##1}}}
\expandafter\def\csname PYGdefault@tok@cp\endcsname{\def\PYGdefault@tc##1{\textcolor[rgb]{0.74,0.48,0.00}{##1}}}
\expandafter\def\csname PYGdefault@tok@k\endcsname{\let\PYGdefault@bf=\textbf\def\PYGdefault@tc##1{\textcolor[rgb]{0.00,0.50,0.00}{##1}}}
\expandafter\def\csname PYGdefault@tok@kp\endcsname{\def\PYGdefault@tc##1{\textcolor[rgb]{0.00,0.50,0.00}{##1}}}
\expandafter\def\csname PYGdefault@tok@kt\endcsname{\def\PYGdefault@tc##1{\textcolor[rgb]{0.69,0.00,0.25}{##1}}}
\expandafter\def\csname PYGdefault@tok@o\endcsname{\def\PYGdefault@tc##1{\textcolor[rgb]{0.40,0.40,0.40}{##1}}}
\expandafter\def\csname PYGdefault@tok@ow\endcsname{\let\PYGdefault@bf=\textbf\def\PYGdefault@tc##1{\textcolor[rgb]{0.67,0.13,1.00}{##1}}}
\expandafter\def\csname PYGdefault@tok@nb\endcsname{\def\PYGdefault@tc##1{\textcolor[rgb]{0.00,0.50,0.00}{##1}}}
\expandafter\def\csname PYGdefault@tok@nf\endcsname{\def\PYGdefault@tc##1{\textcolor[rgb]{0.00,0.00,1.00}{##1}}}
\expandafter\def\csname PYGdefault@tok@nc\endcsname{\let\PYGdefault@bf=\textbf\def\PYGdefault@tc##1{\textcolor[rgb]{0.00,0.00,1.00}{##1}}}
\expandafter\def\csname PYGdefault@tok@nn\endcsname{\let\PYGdefault@bf=\textbf\def\PYGdefault@tc##1{\textcolor[rgb]{0.00,0.00,1.00}{##1}}}
\expandafter\def\csname PYGdefault@tok@ne\endcsname{\let\PYGdefault@bf=\textbf\def\PYGdefault@tc##1{\textcolor[rgb]{0.82,0.25,0.23}{##1}}}
\expandafter\def\csname PYGdefault@tok@nv\endcsname{\def\PYGdefault@tc##1{\textcolor[rgb]{0.10,0.09,0.49}{##1}}}
\expandafter\def\csname PYGdefault@tok@no\endcsname{\def\PYGdefault@tc##1{\textcolor[rgb]{0.53,0.00,0.00}{##1}}}
\expandafter\def\csname PYGdefault@tok@nl\endcsname{\def\PYGdefault@tc##1{\textcolor[rgb]{0.63,0.63,0.00}{##1}}}
\expandafter\def\csname PYGdefault@tok@ni\endcsname{\let\PYGdefault@bf=\textbf\def\PYGdefault@tc##1{\textcolor[rgb]{0.60,0.60,0.60}{##1}}}
\expandafter\def\csname PYGdefault@tok@na\endcsname{\def\PYGdefault@tc##1{\textcolor[rgb]{0.49,0.56,0.16}{##1}}}
\expandafter\def\csname PYGdefault@tok@nt\endcsname{\let\PYGdefault@bf=\textbf\def\PYGdefault@tc##1{\textcolor[rgb]{0.00,0.50,0.00}{##1}}}
\expandafter\def\csname PYGdefault@tok@nd\endcsname{\def\PYGdefault@tc##1{\textcolor[rgb]{0.67,0.13,1.00}{##1}}}
\expandafter\def\csname PYGdefault@tok@s\endcsname{\def\PYGdefault@tc##1{\textcolor[rgb]{0.73,0.13,0.13}{##1}}}
\expandafter\def\csname PYGdefault@tok@sd\endcsname{\let\PYGdefault@it=\textit\def\PYGdefault@tc##1{\textcolor[rgb]{0.73,0.13,0.13}{##1}}}
\expandafter\def\csname PYGdefault@tok@si\endcsname{\let\PYGdefault@bf=\textbf\def\PYGdefault@tc##1{\textcolor[rgb]{0.73,0.40,0.53}{##1}}}
\expandafter\def\csname PYGdefault@tok@se\endcsname{\let\PYGdefault@bf=\textbf\def\PYGdefault@tc##1{\textcolor[rgb]{0.73,0.40,0.13}{##1}}}
\expandafter\def\csname PYGdefault@tok@sr\endcsname{\def\PYGdefault@tc##1{\textcolor[rgb]{0.73,0.40,0.53}{##1}}}
\expandafter\def\csname PYGdefault@tok@ss\endcsname{\def\PYGdefault@tc##1{\textcolor[rgb]{0.10,0.09,0.49}{##1}}}
\expandafter\def\csname PYGdefault@tok@sx\endcsname{\def\PYGdefault@tc##1{\textcolor[rgb]{0.00,0.50,0.00}{##1}}}
\expandafter\def\csname PYGdefault@tok@m\endcsname{\def\PYGdefault@tc##1{\textcolor[rgb]{0.40,0.40,0.40}{##1}}}
\expandafter\def\csname PYGdefault@tok@gh\endcsname{\let\PYGdefault@bf=\textbf\def\PYGdefault@tc##1{\textcolor[rgb]{0.00,0.00,0.50}{##1}}}
\expandafter\def\csname PYGdefault@tok@gu\endcsname{\let\PYGdefault@bf=\textbf\def\PYGdefault@tc##1{\textcolor[rgb]{0.50,0.00,0.50}{##1}}}
\expandafter\def\csname PYGdefault@tok@gd\endcsname{\def\PYGdefault@tc##1{\textcolor[rgb]{0.63,0.00,0.00}{##1}}}
\expandafter\def\csname PYGdefault@tok@gi\endcsname{\def\PYGdefault@tc##1{\textcolor[rgb]{0.00,0.63,0.00}{##1}}}
\expandafter\def\csname PYGdefault@tok@gr\endcsname{\def\PYGdefault@tc##1{\textcolor[rgb]{1.00,0.00,0.00}{##1}}}
\expandafter\def\csname PYGdefault@tok@ge\endcsname{\let\PYGdefault@it=\textit}
\expandafter\def\csname PYGdefault@tok@gs\endcsname{\let\PYGdefault@bf=\textbf}
\expandafter\def\csname PYGdefault@tok@gp\endcsname{\let\PYGdefault@bf=\textbf\def\PYGdefault@tc##1{\textcolor[rgb]{0.00,0.00,0.50}{##1}}}
\expandafter\def\csname PYGdefault@tok@go\endcsname{\def\PYGdefault@tc##1{\textcolor[rgb]{0.53,0.53,0.53}{##1}}}
\expandafter\def\csname PYGdefault@tok@gt\endcsname{\def\PYGdefault@tc##1{\textcolor[rgb]{0.00,0.27,0.87}{##1}}}
\expandafter\def\csname PYGdefault@tok@err\endcsname{\def\PYGdefault@bc##1{\setlength{\fboxsep}{0pt}\fcolorbox[rgb]{1.00,0.00,0.00}{1,1,1}{\strut ##1}}}
\expandafter\def\csname PYGdefault@tok@kc\endcsname{\let\PYGdefault@bf=\textbf\def\PYGdefault@tc##1{\textcolor[rgb]{0.00,0.50,0.00}{##1}}}
\expandafter\def\csname PYGdefault@tok@kd\endcsname{\let\PYGdefault@bf=\textbf\def\PYGdefault@tc##1{\textcolor[rgb]{0.00,0.50,0.00}{##1}}}
\expandafter\def\csname PYGdefault@tok@kn\endcsname{\let\PYGdefault@bf=\textbf\def\PYGdefault@tc##1{\textcolor[rgb]{0.00,0.50,0.00}{##1}}}
\expandafter\def\csname PYGdefault@tok@kr\endcsname{\let\PYGdefault@bf=\textbf\def\PYGdefault@tc##1{\textcolor[rgb]{0.00,0.50,0.00}{##1}}}
\expandafter\def\csname PYGdefault@tok@bp\endcsname{\def\PYGdefault@tc##1{\textcolor[rgb]{0.00,0.50,0.00}{##1}}}
\expandafter\def\csname PYGdefault@tok@fm\endcsname{\def\PYGdefault@tc##1{\textcolor[rgb]{0.00,0.00,1.00}{##1}}}
\expandafter\def\csname PYGdefault@tok@vc\endcsname{\def\PYGdefault@tc##1{\textcolor[rgb]{0.10,0.09,0.49}{##1}}}
\expandafter\def\csname PYGdefault@tok@vg\endcsname{\def\PYGdefault@tc##1{\textcolor[rgb]{0.10,0.09,0.49}{##1}}}
\expandafter\def\csname PYGdefault@tok@vi\endcsname{\def\PYGdefault@tc##1{\textcolor[rgb]{0.10,0.09,0.49}{##1}}}
\expandafter\def\csname PYGdefault@tok@vm\endcsname{\def\PYGdefault@tc##1{\textcolor[rgb]{0.10,0.09,0.49}{##1}}}
\expandafter\def\csname PYGdefault@tok@sa\endcsname{\def\PYGdefault@tc##1{\textcolor[rgb]{0.73,0.13,0.13}{##1}}}
\expandafter\def\csname PYGdefault@tok@sb\endcsname{\def\PYGdefault@tc##1{\textcolor[rgb]{0.73,0.13,0.13}{##1}}}
\expandafter\def\csname PYGdefault@tok@sc\endcsname{\def\PYGdefault@tc##1{\textcolor[rgb]{0.73,0.13,0.13}{##1}}}
\expandafter\def\csname PYGdefault@tok@dl\endcsname{\def\PYGdefault@tc##1{\textcolor[rgb]{0.73,0.13,0.13}{##1}}}
\expandafter\def\csname PYGdefault@tok@s2\endcsname{\def\PYGdefault@tc##1{\textcolor[rgb]{0.73,0.13,0.13}{##1}}}
\expandafter\def\csname PYGdefault@tok@sh\endcsname{\def\PYGdefault@tc##1{\textcolor[rgb]{0.73,0.13,0.13}{##1}}}
\expandafter\def\csname PYGdefault@tok@s1\endcsname{\def\PYGdefault@tc##1{\textcolor[rgb]{0.73,0.13,0.13}{##1}}}
\expandafter\def\csname PYGdefault@tok@mb\endcsname{\def\PYGdefault@tc##1{\textcolor[rgb]{0.40,0.40,0.40}{##1}}}
\expandafter\def\csname PYGdefault@tok@mf\endcsname{\def\PYGdefault@tc##1{\textcolor[rgb]{0.40,0.40,0.40}{##1}}}
\expandafter\def\csname PYGdefault@tok@mh\endcsname{\def\PYGdefault@tc##1{\textcolor[rgb]{0.40,0.40,0.40}{##1}}}
\expandafter\def\csname PYGdefault@tok@mi\endcsname{\def\PYGdefault@tc##1{\textcolor[rgb]{0.40,0.40,0.40}{##1}}}
\expandafter\def\csname PYGdefault@tok@il\endcsname{\def\PYGdefault@tc##1{\textcolor[rgb]{0.40,0.40,0.40}{##1}}}
\expandafter\def\csname PYGdefault@tok@mo\endcsname{\def\PYGdefault@tc##1{\textcolor[rgb]{0.40,0.40,0.40}{##1}}}
\expandafter\def\csname PYGdefault@tok@ch\endcsname{\let\PYGdefault@it=\textit\def\PYGdefault@tc##1{\textcolor[rgb]{0.25,0.50,0.50}{##1}}}
\expandafter\def\csname PYGdefault@tok@cm\endcsname{\let\PYGdefault@it=\textit\def\PYGdefault@tc##1{\textcolor[rgb]{0.25,0.50,0.50}{##1}}}
\expandafter\def\csname PYGdefault@tok@cpf\endcsname{\let\PYGdefault@it=\textit\def\PYGdefault@tc##1{\textcolor[rgb]{0.25,0.50,0.50}{##1}}}
\expandafter\def\csname PYGdefault@tok@c1\endcsname{\let\PYGdefault@it=\textit\def\PYGdefault@tc##1{\textcolor[rgb]{0.25,0.50,0.50}{##1}}}
\expandafter\def\csname PYGdefault@tok@cs\endcsname{\let\PYGdefault@it=\textit\def\PYGdefault@tc##1{\textcolor[rgb]{0.25,0.50,0.50}{##1}}}

\makeatother

\makeatletter
\def\PYG@reset{\let\PYG@it=\relax \let\PYG@bf=\relax%
    \let\PYG@ul=\relax \let\PYG@tc=\relax%
    \let\PYG@bc=\relax \let\PYG@ff=\relax}
\def\PYG@tok#1{\csname PYG@tok@#1\endcsname}
\def\PYG@toks#1+{\ifx\relax#1\empty\else%
    \PYG@tok{#1}\expandafter\PYG@toks\fi}
\def\PYG@do#1{\PYG@bc{\PYG@tc{\PYG@ul{%
    \PYG@it{\PYG@bf{\PYG@ff{#1}}}}}}}
\def\PYG#1#2{\PYG@reset\PYG@toks#1+\relax+\PYG@do{#2}}

\expandafter\def\csname PYG@tok@w\endcsname{\def\PYG@tc##1{\textcolor[rgb]{0.73,0.73,0.73}{##1}}}
\expandafter\def\csname PYG@tok@c\endcsname{\let\PYG@it=\textit\def\PYG@tc##1{\textcolor[rgb]{0.25,0.50,0.50}{##1}}}
\expandafter\def\csname PYG@tok@cp\endcsname{\def\PYG@tc##1{\textcolor[rgb]{0.74,0.48,0.00}{##1}}}
\expandafter\def\csname PYG@tok@k\endcsname{\let\PYG@bf=\textbf\def\PYG@tc##1{\textcolor[rgb]{0.00,0.50,0.00}{##1}}}
\expandafter\def\csname PYG@tok@kp\endcsname{\def\PYG@tc##1{\textcolor[rgb]{0.00,0.50,0.00}{##1}}}
\expandafter\def\csname PYG@tok@kt\endcsname{\def\PYG@tc##1{\textcolor[rgb]{0.69,0.00,0.25}{##1}}}
\expandafter\def\csname PYG@tok@o\endcsname{\def\PYG@tc##1{\textcolor[rgb]{0.40,0.40,0.40}{##1}}}
\expandafter\def\csname PYG@tok@ow\endcsname{\let\PYG@bf=\textbf\def\PYG@tc##1{\textcolor[rgb]{0.67,0.13,1.00}{##1}}}
\expandafter\def\csname PYG@tok@nb\endcsname{\def\PYG@tc##1{\textcolor[rgb]{0.00,0.50,0.00}{##1}}}
\expandafter\def\csname PYG@tok@nf\endcsname{\def\PYG@tc##1{\textcolor[rgb]{0.00,0.00,1.00}{##1}}}
\expandafter\def\csname PYG@tok@nc\endcsname{\let\PYG@bf=\textbf\def\PYG@tc##1{\textcolor[rgb]{0.00,0.00,1.00}{##1}}}
\expandafter\def\csname PYG@tok@nn\endcsname{\let\PYG@bf=\textbf\def\PYG@tc##1{\textcolor[rgb]{0.00,0.00,1.00}{##1}}}
\expandafter\def\csname PYG@tok@ne\endcsname{\let\PYG@bf=\textbf\def\PYG@tc##1{\textcolor[rgb]{0.82,0.25,0.23}{##1}}}
\expandafter\def\csname PYG@tok@nv\endcsname{\def\PYG@tc##1{\textcolor[rgb]{0.10,0.09,0.49}{##1}}}
\expandafter\def\csname PYG@tok@no\endcsname{\def\PYG@tc##1{\textcolor[rgb]{0.53,0.00,0.00}{##1}}}
\expandafter\def\csname PYG@tok@nl\endcsname{\def\PYG@tc##1{\textcolor[rgb]{0.63,0.63,0.00}{##1}}}
\expandafter\def\csname PYG@tok@ni\endcsname{\let\PYG@bf=\textbf\def\PYG@tc##1{\textcolor[rgb]{0.60,0.60,0.60}{##1}}}
\expandafter\def\csname PYG@tok@na\endcsname{\def\PYG@tc##1{\textcolor[rgb]{0.49,0.56,0.16}{##1}}}
\expandafter\def\csname PYG@tok@nt\endcsname{\let\PYG@bf=\textbf\def\PYG@tc##1{\textcolor[rgb]{0.00,0.50,0.00}{##1}}}
\expandafter\def\csname PYG@tok@nd\endcsname{\def\PYG@tc##1{\textcolor[rgb]{0.67,0.13,1.00}{##1}}}
\expandafter\def\csname PYG@tok@s\endcsname{\def\PYG@tc##1{\textcolor[rgb]{0.73,0.13,0.13}{##1}}}
\expandafter\def\csname PYG@tok@sd\endcsname{\let\PYG@it=\textit\def\PYG@tc##1{\textcolor[rgb]{0.73,0.13,0.13}{##1}}}
\expandafter\def\csname PYG@tok@si\endcsname{\let\PYG@bf=\textbf\def\PYG@tc##1{\textcolor[rgb]{0.73,0.40,0.53}{##1}}}
\expandafter\def\csname PYG@tok@se\endcsname{\let\PYG@bf=\textbf\def\PYG@tc##1{\textcolor[rgb]{0.73,0.40,0.13}{##1}}}
\expandafter\def\csname PYG@tok@sr\endcsname{\def\PYG@tc##1{\textcolor[rgb]{0.73,0.40,0.53}{##1}}}
\expandafter\def\csname PYG@tok@ss\endcsname{\def\PYG@tc##1{\textcolor[rgb]{0.10,0.09,0.49}{##1}}}
\expandafter\def\csname PYG@tok@sx\endcsname{\def\PYG@tc##1{\textcolor[rgb]{0.00,0.50,0.00}{##1}}}
\expandafter\def\csname PYG@tok@m\endcsname{\def\PYG@tc##1{\textcolor[rgb]{0.40,0.40,0.40}{##1}}}
\expandafter\def\csname PYG@tok@gh\endcsname{\let\PYG@bf=\textbf\def\PYG@tc##1{\textcolor[rgb]{0.00,0.00,0.50}{##1}}}
\expandafter\def\csname PYG@tok@gu\endcsname{\let\PYG@bf=\textbf\def\PYG@tc##1{\textcolor[rgb]{0.50,0.00,0.50}{##1}}}
\expandafter\def\csname PYG@tok@gd\endcsname{\def\PYG@tc##1{\textcolor[rgb]{0.63,0.00,0.00}{##1}}}
\expandafter\def\csname PYG@tok@gi\endcsname{\def\PYG@tc##1{\textcolor[rgb]{0.00,0.63,0.00}{##1}}}
\expandafter\def\csname PYG@tok@gr\endcsname{\def\PYG@tc##1{\textcolor[rgb]{1.00,0.00,0.00}{##1}}}
\expandafter\def\csname PYG@tok@ge\endcsname{\let\PYG@it=\textit}
\expandafter\def\csname PYG@tok@gs\endcsname{\let\PYG@bf=\textbf}
\expandafter\def\csname PYG@tok@gp\endcsname{\let\PYG@bf=\textbf\def\PYG@tc##1{\textcolor[rgb]{0.00,0.00,0.50}{##1}}}
\expandafter\def\csname PYG@tok@go\endcsname{\def\PYG@tc##1{\textcolor[rgb]{0.53,0.53,0.53}{##1}}}
\expandafter\def\csname PYG@tok@gt\endcsname{\def\PYG@tc##1{\textcolor[rgb]{0.00,0.27,0.87}{##1}}}
\expandafter\def\csname PYG@tok@err\endcsname{\def\PYG@bc##1{\setlength{\fboxsep}{0pt}\fcolorbox[rgb]{1.00,0.00,0.00}{1,1,1}{\strut ##1}}}
\expandafter\def\csname PYG@tok@kc\endcsname{\let\PYG@bf=\textbf\def\PYG@tc##1{\textcolor[rgb]{0.00,0.50,0.00}{##1}}}
\expandafter\def\csname PYG@tok@kd\endcsname{\let\PYG@bf=\textbf\def\PYG@tc##1{\textcolor[rgb]{0.00,0.50,0.00}{##1}}}
\expandafter\def\csname PYG@tok@kn\endcsname{\let\PYG@bf=\textbf\def\PYG@tc##1{\textcolor[rgb]{0.00,0.50,0.00}{##1}}}
\expandafter\def\csname PYG@tok@kr\endcsname{\let\PYG@bf=\textbf\def\PYG@tc##1{\textcolor[rgb]{0.00,0.50,0.00}{##1}}}
\expandafter\def\csname PYG@tok@bp\endcsname{\def\PYG@tc##1{\textcolor[rgb]{0.00,0.50,0.00}{##1}}}
\expandafter\def\csname PYG@tok@fm\endcsname{\def\PYG@tc##1{\textcolor[rgb]{0.00,0.00,1.00}{##1}}}
\expandafter\def\csname PYG@tok@vc\endcsname{\def\PYG@tc##1{\textcolor[rgb]{0.10,0.09,0.49}{##1}}}
\expandafter\def\csname PYG@tok@vg\endcsname{\def\PYG@tc##1{\textcolor[rgb]{0.10,0.09,0.49}{##1}}}
\expandafter\def\csname PYG@tok@vi\endcsname{\def\PYG@tc##1{\textcolor[rgb]{0.10,0.09,0.49}{##1}}}
\expandafter\def\csname PYG@tok@vm\endcsname{\def\PYG@tc##1{\textcolor[rgb]{0.10,0.09,0.49}{##1}}}
\expandafter\def\csname PYG@tok@sa\endcsname{\def\PYG@tc##1{\textcolor[rgb]{0.73,0.13,0.13}{##1}}}
\expandafter\def\csname PYG@tok@sb\endcsname{\def\PYG@tc##1{\textcolor[rgb]{0.73,0.13,0.13}{##1}}}
\expandafter\def\csname PYG@tok@sc\endcsname{\def\PYG@tc##1{\textcolor[rgb]{0.73,0.13,0.13}{##1}}}
\expandafter\def\csname PYG@tok@dl\endcsname{\def\PYG@tc##1{\textcolor[rgb]{0.73,0.13,0.13}{##1}}}
\expandafter\def\csname PYG@tok@s2\endcsname{\def\PYG@tc##1{\textcolor[rgb]{0.73,0.13,0.13}{##1}}}
\expandafter\def\csname PYG@tok@sh\endcsname{\def\PYG@tc##1{\textcolor[rgb]{0.73,0.13,0.13}{##1}}}
\expandafter\def\csname PYG@tok@s1\endcsname{\def\PYG@tc##1{\textcolor[rgb]{0.73,0.13,0.13}{##1}}}
\expandafter\def\csname PYG@tok@mb\endcsname{\def\PYG@tc##1{\textcolor[rgb]{0.40,0.40,0.40}{##1}}}
\expandafter\def\csname PYG@tok@mf\endcsname{\def\PYG@tc##1{\textcolor[rgb]{0.40,0.40,0.40}{##1}}}
\expandafter\def\csname PYG@tok@mh\endcsname{\def\PYG@tc##1{\textcolor[rgb]{0.40,0.40,0.40}{##1}}}
\expandafter\def\csname PYG@tok@mi\endcsname{\def\PYG@tc##1{\textcolor[rgb]{0.40,0.40,0.40}{##1}}}
\expandafter\def\csname PYG@tok@il\endcsname{\def\PYG@tc##1{\textcolor[rgb]{0.40,0.40,0.40}{##1}}}
\expandafter\def\csname PYG@tok@mo\endcsname{\def\PYG@tc##1{\textcolor[rgb]{0.40,0.40,0.40}{##1}}}
\expandafter\def\csname PYG@tok@ch\endcsname{\let\PYG@it=\textit\def\PYG@tc##1{\textcolor[rgb]{0.25,0.50,0.50}{##1}}}
\expandafter\def\csname PYG@tok@cm\endcsname{\let\PYG@it=\textit\def\PYG@tc##1{\textcolor[rgb]{0.25,0.50,0.50}{##1}}}
\expandafter\def\csname PYG@tok@cpf\endcsname{\let\PYG@it=\textit\def\PYG@tc##1{\textcolor[rgb]{0.25,0.50,0.50}{##1}}}
\expandafter\def\csname PYG@tok@c1\endcsname{\let\PYG@it=\textit\def\PYG@tc##1{\textcolor[rgb]{0.25,0.50,0.50}{##1}}}
\expandafter\def\csname PYG@tok@cs\endcsname{\let\PYG@it=\textit\def\PYG@tc##1{\textcolor[rgb]{0.25,0.50,0.50}{##1}}}

\def\PYGZus{\char`\_}

\def\PYGZsh{\char`\#}

\def\PYGZhy{\char`\-}
\def\PYGZsq{\char`\'}

\makeatother

\usepackage{float}
\usepackage{graphicx}
\usepackage{fontawesome}
\usepackage{tikz}
\usepackage{tikz-3dplot-circleofsphere}
\usepackage{tikz-3dplot}
\usepackage{bm}
\usetikzlibrary{bayesnet}
\usepackage{amsmath}
\usepackage{widetext}
\usepackage{lineno}
\usepackage{multirow}
\usepackage{subfigure}

\usepackage{scalerel}
\usepackage[]{xcolor}
\definecolor{citecolor}{HTML}{195D95}
\definecolor{BrickRed}{HTML}{C41010}

\definecolor{green}{HTML}{26E065}
\definecolor{dark}{HTML}{930A85}
\definecolor{mid}{HTML}{c04bb4}
\definecolor{light}{HTML}{d692cf}

\definecolor{globe}{RGB}{138,138,138}
\definecolor{det1color}{HTML}{26E065}

\usepackage[colorlinks=true,linkcolor=mid,citecolor=dark,urlcolor=green]{hyperref}

\usepackage{amsmath,amssymb,latexsym}

\def\r{4}
\tdplotsetmaincoords{91}{50}

\def\pgrb{\vec{p}_{\scaleto{\mathrm{GRB}}{3pt}}}
  
\newcommand{\pos}[1]{\vec{p}_{#1}}

\begin{document}
\title{nazgul: A statistical approach to gamma-ray burst localization} 
\subtitle{Triangulation via non-stationary time-series models}

\author{J. Michael Burgess \inst{1} 
  \and Ewan Cameron\inst{2}
  \and Dmitry Svinkin\inst{3}
  \and Jochen Greiner\inst{1}
}

\institute{Max-Planck-Institut fur extraterrestrische Physik, Giessenbachstrasse 1, D-85748 Garching, Germany \\
  \email{jburgess@mpe.mpg.de}\label{mpe}
  \and Curtin University, Kent Street, Bentley, Perth, 6102 WA, Australia\label{ec}
  \and Ioffe Institute, 26 Politekhnicheskaya, St Petersburg, 194021, Russia
}
\date{}

\label{firstpage}

\abstract {Gamma-ray bursts can be located via arrival time signal
  triangulation using gamma-ray detectors in orbit throughout the
  solar system. The classical approach based on cross-correlations of
  binned light curves ignores the Poisson nature of the time-series
  data, and is unable to model the full complexity of the problem.}
{To present a statistically proper and robust GRB timing/triangulation
  algorithm as a modern update to the original procedures used for the
  Interplanetary Network (IPN).}  {A hierarchical Bayesian forward
  model for the unknown temporal signal evolution is learned via
  random Fourier features (RFF) and fitted to each detector's
  time-series data with time-differences that correspond to GRB's
  position on the sky via the appropriate Poisson likelihood.} {Our
  novel method can robustly estimate the position of a GRB as verified
  via simulations. The uncertainties generated by the method are
  robust and in many cases more precise compared to the classical
  method. Thus, we have a method that can become a valuable tool for
  gravitational wave follow-up. All software and analysis scripts are
  made publicly available
  \faGithub\href{https://github.com/grburgess/nazgul}{~here} for the
  purpose of replication.} {}

\keywords{ (stars:) gamma ray bursts -- methods: data analysis -- methods: statistical}

\maketitle

\section{Introduction}
Estimating the precise location on the celestial sphere from which a
detected gamma-ray burst (GRB) has been received---a process known as
`localization'---has become a critical component of a multi-national,
multi-mission campaign to alert astronomers to observational follow-up
opportunities regarding associated transients
\citep[e.g.][]{Hurley:2019aa,Kozlova:2016aa,Hurley:2000ab},
gravitational waves \citep{Hurley:2016aa}, and neutrinos
\citep{Aartsen:2017aa}.  All-sky survey missions optimised for
temporal and spectral signal resolution, such as the Fermi Gamma-Ray
Burst monitor (GBM; \citealt{Meegan:2009aa}), Konus
\citep{Aptekar:1995aa} or INTEGRAL SPI-ACS \citep{Vedrenne:2003aa},
have a high sensitivity for detection of an incoming GRB, but either
offer no spatial information or struggle to achieve precise
localizations. Fermi GBM, for example, performs a statistical
localization based on model fitting to the relative signal strengths
received through its twelve directional detectors, however, an
effective systematic uncertainty of 3.7 degrees
\citep{Connaughton:2015aa} equates to a search area
orders of magnitude larger than the field of view of key follow up
instruments.  An alternative and complementary localization approach
is offered by the Interplanetary Network (IPN)
\citep{Hurley:1999aa,Palshin:2013aa} of gamma-ray detecting
satellites, in particular those beyond Earth-centered orbits, which
enable the use of a triangulation-based positioning strategy
harnessing the estimated differences between arrival times of a GRB
signal at each detector.  For the combination of independent estimates
from these two approaches to properly refine the search area for
follow up, it is essential that the uncertainties of both are well
calibrated
\citep{Hurley:2013aa}.  

The classical IPN (hereafter cIPN) algorithm identifies time-delays in
the light curves of gamma-ray detectors in its network via the
cross-correlation of background subtracted time-series (see, e.g.,
\citealt{Palshin:2013aa}).  A point estimate for the time delay is
computed for the case of two GRB signals with equal bin size in the
time domain by finding the discrete bin shift that yields the minimum
``reduced-$\chi^2$''; the latter being a summary statistic representing
the log-likelihood of the observed signal difference at a given time
shift under a Normal approximation to the Poisson count distribution.
Confidence intervals (possibly asymmetric) on this time delay are
computed via the likelihood ratio test \citep{Wilks:1938aa} and used
to create annular regions on the sky which indicate the probabilistic
location of the GRB (see Section \ref{sec:concept} for further
details).  While this method is conceptually and algorithmically
simple it based on dual asymptotic approximations disfavouring the low
count and/or small bin regime. Furthermore, this method cannot be
naturally extended to handle systematic complexities of the
observational process, such as differences in energy dispersion
between detectors or time-varying backgrounds.

To improve upon the cIPN, we propose a new IPN algorithm we call {\tt
  nazgul}\footnote{This refers to the Tolkien characters in search of
  a magic ring, but in this case, annuli.} built on a hierarchical,
forward modeling approach to likelihood-based inference. We will
demonstrate through the analysis of mock data sets generated via
realistic simulations that the {\tt nazgul} algorithm produces precise
statistical localizations with well-calibrated uncertainties, even in
challenging settings where the cIPN algorithm does not.  As much of
the data necessary for IPN localizations of real observed sources is
not (at present) publicly available we restrict our investigation to
the simulated regime and focus on the methodological development. We
plan a future study with real data to assess its viability and
systematics. However, we explicitly make available all the required
software \faGithub\href{https://github.com/grburgess/nazgul}{~here} to
reproduce our results and to assist in further development of the
approach with the hope that {\tt nazgul} can become a valuable tool
for the GRB localization community.

\section{Concept}
\label{sec:concept}
\begin{figure}[h]
  \centering
 \begin{tikzpicture}[tdplot_main_coords]
\coordinate (O) at (0,0,0);
 \foreach \a in {-65,-40,...,60}
  {\tdplotCsDrawLatCircle[very thin,black!30]{\r}{\a}}
 \foreach \a in {0,45,...,315} 
  {\tdplotCsDrawLonCircle[very thin,black!30]{\r}{\a}}

\draw [ball color=globe,very thin,tdplot_screen_coords, opacity=0.75 ] (0,0,0) circle (\r);
\draw[-stealth,color=dark] (O) -- (0.345646175253801,0.29003157817330183,0.039475675592239245) node[below] {$p_1$};
\draw[-stealth,color=dark] (O) -- (-3.084721927994411,1.1227469627782176,-0.5788272588897677) node[above] {$p_2$};
 \draw [ball color=blue,very thin,tdplot_screen_coords, opacity=0.6 ] (0,0,0) circle (0.041953663118346904);
\tdplotCsDrawCircle[color=det1color,thick,tdplotCsFill/.style={opacity=0.00}]{\r}{166.35548003090486}{99.93497667546379}{-47.65160072722878}
\draw[-stealth,color=det1color] (O) -- (3.8288215840350883,-0.9294392152592506,0.6901217448865852) node[right] {$p_1 - p_2$};
\tdplotdefinepoints (0,0,0)(3.8288215840350883,-0.9294392152592506,0.6901217448865852)(2.8793852415718173,1.0480105209175399,2.5711504387461566) 
 \tdplotdrawpolytopearc[dark,thin]{1}{anchor=west} {$\theta$}
\draw[dashed,->,color=dark] (O) -- (2.8793852415718173,1.0480105209175399,2.5711504387461566) node[left] {$p_{grb}$};
 
 \end{tikzpicture}  
\caption{Illustrative example of the IPN geometry for two
  detectors. One is in low-Earth orbit, and the other in an orbit
  around the forest moon of Endor.}
  \label{fig:thatsnomoon}
\end{figure}
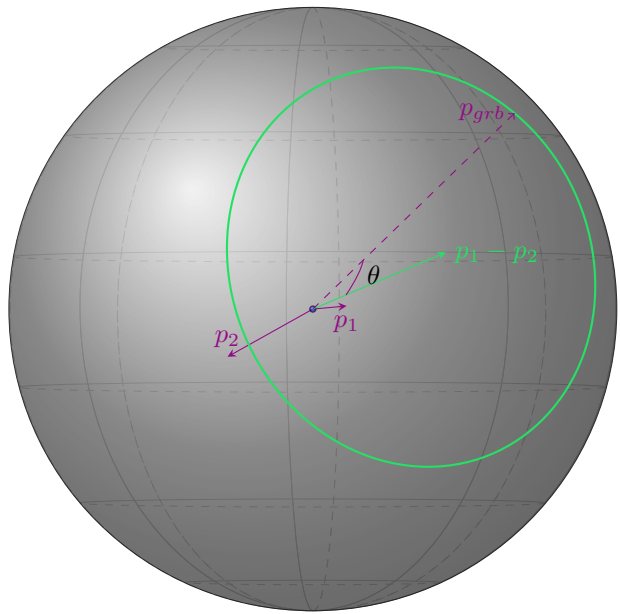

Consider a GRB with a position vector $\pgrb$
detected by two GRB detectors each with position vectors $\vec{p}_1$
and $\vec{p}_2$ respectively. The time-difference of the arrival of
the signal at each detector is given by 

\begin{equation}
  \label{eq:dt}
  \Delta t(\pgrb,\pos{1},\pos{2}) = \frac{\pgrb \cdot (\pos{1} - \pos{2})}{c} \text{.}
\end{equation}
\noindent
Using this time difference, one can construct an annulus on the sky on
which the 2D position of the GRB lies. This annulus is computed by
inverting Eq. \ref{eq:dt} such that

\begin{equation}
  \label{eq:angle}
  \theta = \cos^{-1}\left(\frac{c \Delta t }{\| \pos{1} - \pos{2} \|} \right)
\end{equation}
\noindent
is the opening angle of the annulus on the sky with a center in the
direction of $\pos{1} - \pos{2}$ as shown in Figure
\ref{fig:thatsnomoon}. The error in the time measurement is typically
taken as the sole error, and transformed into a `width' of the
annulus. Additional detectors can be added to the problem resulting in
$N_{\mathrm{det}} (N_{\mathrm{det}}-1)/2$ unique baselines from which
further intersecting annuli can be derived. Thus, the problem of
localizing a GRB, given accurate satellite/detector position and
accurate satellite clocks is reduced to that of accurately determining
$\Delta t$ of each detector baseline.

\begin{figure}[h]
  \centering
  \includegraphics[]{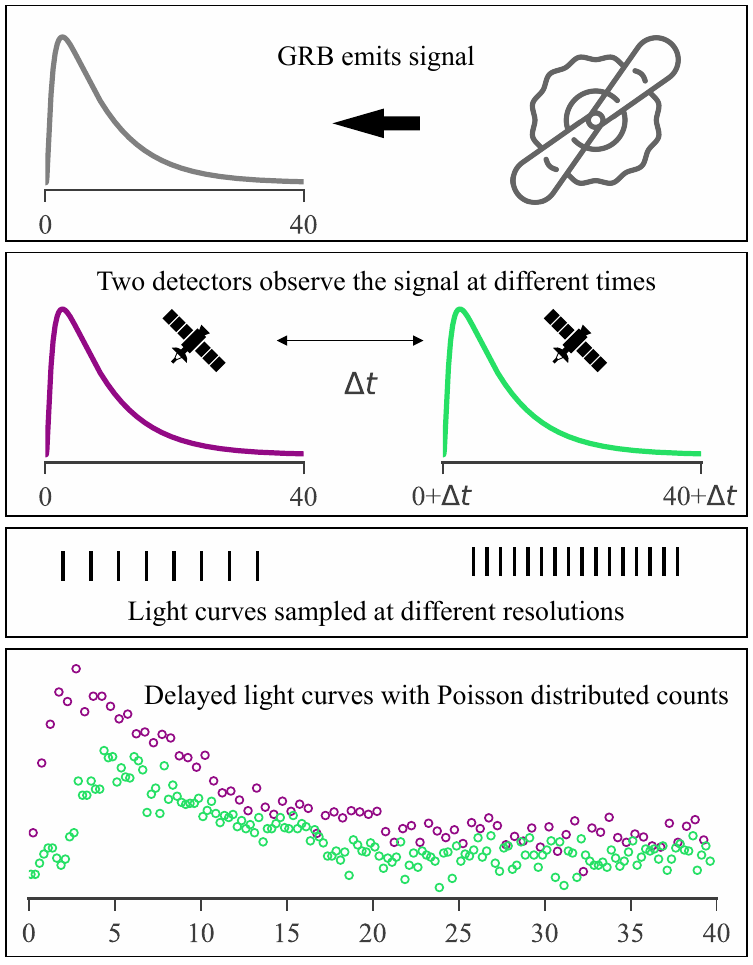}
  \caption{An illustrative sketch of the problem as well as the
    framework for our forward model. A GRB emits a signal which is
    detected at two different times in detectors at different
    locations. These signals are translated into data sampled with
    different temporal resolutions and different detector effective
    areas resulting in two light curves which are delayed with respect
    to each other. }
  \label{fig:toy}
\end{figure}

\section{Methodology}

Here we propose to estimate the inter-satellite time delay(s) of each
GRB signal using a Bayesian (i.e., prior-regularised,
likelihood-based) inference procedure built around a forward model
designed to acknowledge the inherent stochasticity of the process by
which the observational data are generated as shown in Figure
\ref{fig:toy}.  To this end we suppose that the GRB emits a stream of
photons distributed as the realisation of a Poisson process indexed by
time with a time-varying intensity such that each detector, $d_i$ for
$i\in {1,\ldots,N_{\mathrm{det}}}$, receives a portion of these photons along
its unique line of sight.  The effective temporal intensity function
of each received signal, $S_i(t)$, is thus identical modulo its
specific source-to-detector time delay, which for convenience we
define with respect to the $i=1$ labeled detector as $\Delta t_i$ with
$\Delta t_1 = 0$, hence $S_i(t) \equiv S(t+\Delta t_i)$.  We further suppose that
each detector receives an independent background signal also
distributed as a Poisson process with an intensity, $B_i(t)$,
combining additively with the GRB signal towards a Poisson process
with total intensity function, $C_i(t) = S_i(t)+B_i(t)$.

Due to engineering constraints, different detectors can have different
energy-dependent effective areas, thus breaking the above assumption
that the observed signal will be exactly matched in structure between
the detectors. This issue can be addressed within the hierarchical
modelling framework by folding the signal model through each
detector's response function, either fitting for or assuming a
spectral signal shape.  For simplicity of exposition and to reduce the
computational burden we will assume here that neither detector suffers
energy dispersion, but will allow for the effective area of each to
vary by a scale factor, $A_i$, such that now
$C_i(t) = A_i S_i(t) + B_i(t)$.  Also note that rather than fitting
for a time difference directly, we can parameterize the time
difference as a function of $\pgrb$ therefore allowing only time
differences that would properly place the GRB's location in the sky;
that is,
\begin{equation*}
\Delta t_{i}(\pgrb,\pos{1},\pos{i}) = \frac{\pgrb \cdot (\pos{1} - \pos{i})}{c}
\end{equation*}
\noindent 
therefore excluding time differences longer than
$\frac{(\pos{1} - \pos{i})}{c}$.

\subsection{Likelihood}
At the top level of our hierarchical model is the likelihood function, $\mathcal{L}(\cdot)$,
for the observed process of photons arriving at our set of detectors,
$\{c_i(t)\}_{i=1,\ldots,N_\mathrm{det}}$,
\begin{equation*}
    \mathcal{L}(\{c_i(t)\}_{i=1,\ldots,N_\mathrm{det}}|\{C_i(t)\}_{i=1,\ldots,N_\mathrm{det}})
    =\prod_{i=1,\ldots,N_\mathrm{det}} \mathcal{L}(c_i(t)|C_i(t)).
\end{equation*}
Here we use the notational form,
$\mathcal{L}(\mathrm{data}|\mathrm{parameters})$, defining the
likelihood as the probability density of the observed data conditional
on a complete set of parameters characterising the assumed
distribution of the sampling process.  The second step is a
factorization into product form since the realized Poisson process of
photons arriving at each detector is conditionally independent once
the mean intensity function, $C_i(t)$, for each is specified.  While
the signal received in each is the realization of continuous time
Poisson process (i.e., a randomly sized set of photons with unique
arrival times), a discretization into counts per small equal unit of
time is imposed by the detector read out.  Hence, we write the
likelihood in each as the product of ordinary Poisson distributions,
$\mathcal{P}(\mathrm{mean\ expected\ count})[\mathrm{observed\
  count}]$, for the photon count over bins of width, $\delta t$, in the
observing window from time, $t_0$, to time,
$t_0+\delta t\times N_\mathrm{bins}$, thus
\begin{equation*}
    \mathcal{L}(c_i(t)|C_i(t)) = \prod_{j=1,\dots,N_\mathrm{bins}} \mathcal{P}\Big(\int_{t_0+(j-1)\delta t}^{t_0+j\delta t} dt^\prime C_i\big(t^{\prime}\big)\Big)\Big[ c_i(t)\Big].
  \end{equation*}
  
Note that for later computational convenience we treat the signal
  model here as if constant over the width of each temporal bin,
  taking the value at the mid-point of each bin as our reference.

\subsection{GRB signal model}
From the forward modelling perspective, the challenge now is to
construct an appropriate probabilistic representation of the expected
distribution of latent (noise-free) signal shapes, $S(t)$, that may
arise from observable GRB events.  Without a guiding physical model
for the evolution of GRB light curves and with respect to the sheer
diversity of shapes on record over the thousands of GRBs detected
to-date, we follow here a Bayesian non-parametric approach
\citep{Hjort:2010aa}.  That is, we adopt a rich stochastic process
prior as the hypothetical distribution of possible latent signal
shapes, supposing that the latent signal corresponding to any given
GRB is a single realization from this process. A convenient and
well-studied choice for this purpose is the Gaussian process (GP),
which prescribes a multivariate Normal distribution over any index set
of observed time points and is characterised by a mean function and a
covariance function pair \citep{Williams:1996aa}\footnote{A wonderful
  introduction to GPs can be found at \href{https://betanalpha.github.io/assets/case\_studies/gp\_part1/part1.html}{https://betanalpha.github.io/assets/case\_studies/gp\_part1/part1.html}}. Astronomical
applications of Gaussian process-based modelling include the
correction of instrumental systematics in stellar light curves to
assist exoplanet detection \citep{Aigrain:2016aa}, reverberation
mapping of accretion disks around black holes with X-ray time series
\citep{Wilkins:2019aa} and modeling the variability in lensed quasars
to estimate cosmological parameters \citep{Suyu:2017aa}. As in the
generalized linear model setting (where the exponential and logarithm
are used as the transform and its inverse to map an unconstrained
linear regression prediction on $\mathbb{R}$ to the $\mathbb{R}^+$
constrained support of the Poisson or negative binomial expectation;
\citealt{Hilbe:2017aa}), we apply an exponentiation to the Gaussian
process within our hierarchical model, such that in fact it is the
logarithm of the signal that is assumed to follow this distribution,
i.e., $\log S(\cdot) \sim \mathrm{GP}$.

A straight-forward approach to Gaussian process modelling (as followed
in the astronomical examples noted above) is to build a valid mean
function and covariance function pair by choosing a constant for the
former (i.e., $\mu(t)=c$) and a stationary, kernel-based prescription
for the latter, \begin{equation*}\mathrm{cov}(t_m,t_n) =
  K(|t_m-t_n|).\end{equation*} Here the kernel function, $K(\cdot)$, may
be composed of one or more (in summation) standard analytic forms
known to correspond to valid covariance functions (in particular, to
be non-negative definite).  For instance, the squared-exponential
kernel, defined as

\begin{equation}\nonumber
  K(\theta)=\sigma^2\exp(-\frac{1}{2}b^2\theta^2),
\end{equation}
\noindent
with $\sigma$ an amplitude scale parameter and $b$ an inverse length-scale
(or `band-width') parameter.  The term `stationary' refers to the
observation that such a kernel function takes only the time delay (or,
more generally, distance) between points as its argument, such that
the degree of covariance between points a fixed lag apart is imagined
constant over the entire domain of the latent signal.  Computing the
likelihood of a realisation of the latent signal at $N$ distinct
observation times requires construction and inversion of the full
$N \times N$ covariance matrix, with a cost that naively scales in
proportion to $N^3$. Computational efficiency gains to scale of order
$N \log^2 N$ can, however, be made with sophisticated inversion
strategies for some of these commonly-used kernels
\citep{Ambikasaran:2015aa}.

In this analysis we follow an alternative approach to approximate a
non-stationary kernel using random Fourier features (RFFs;
\citealt{Rahimi:2008aa}).  The RFF approximation makes use of
Bochner's theorem \citep{Bochner:1934aa} which describes a
correspondence between each normalised, continuous, positive definite
kernel on the real line and a probability measure in the spectral
space of its Fourier transform.  For example, the spectral measure,
$\pi_K(\omega)$, corresponding to a normalized version of the squared
exponential kernel defined above is the zero-mean Normal distribution
with standard deviation equal to the inverse length-scale, $b$, i.e.,
$\pi_K(\omega) = \mathcal{N}(0,b)[\omega]$.  Once the corresponding spectral
measure is identified for a given kernel, a RFF approximation may
begin by drawing $k$ (typically $k<N$) frequencies,
$\omega^{(i)=1,\ldots,k} \sim \pi_K(\omega)$ to form a so-called `random feature matrix',
$\vec{\phi}$, with $N$ rows (one per observed time point) of $k$ sine and
$k$ cosine terms in the columns, i.e.,
\begin{widetext}
\begin{equation*}
  \vec{\phi} = \Big[ \sin(2\pi \omega^{(1)} \vec{t}), \ldots,  \sin(2\pi \omega^{(k)} \vec{t}), \cos(2\pi \omega^{(1)} \vec{t}), \ldots,  \cos(2\pi \omega^{(k)} \vec{t})\Big].
\end{equation*}
\end{widetext}
\noindent
The RFF version of the GP prior on the latent signal may then be
constructed as a regularised linear regression of features as
\begin{equation*}
  \log S(\vec{t}) = \frac{\sigma}{\sqrt{k}} \vec{\phi \beta} \mathrm{\ with\ }\vec{\beta}:\ \vec{\beta}_{m=1,\ldots,2k} \sim \mathcal{N}(0,1).
\end{equation*}
\noindent
That is, $\vec{\beta}$ is formed as a length $2k$ column vector of elements
with independent standard Normal distributions.  For reference, the
corresponding covariance matrix may be computed as
$C = \frac{\sigma^2}{k}\vec{\phi}\vec{\phi}^\prime$, which is similar to the strategy
employed in some X-ray timeseries analyses \citep{Zoghbi:2012aa}.  An
accessible introduction to the random Fourier feature approximation,
albeit from the perspective of spatial Gaussian random fields rather
than timeseries, is given by \citet{Milton:2019aa}.

To allow greater flexibility in our representation of the latent GRB
signal we take an extension of the random Fourier features method to
define a non-stationary kernel.  As described in \citet{Ton:2018aa}
(and see also \citealt{Remes:2017aa}), Yaglom's generalisation of
Bochner's theorem \citep{Yaglom:1987aa} can be used to show that
spectral measures on product spaces of frequencies give rise to
non-stationary kernels.  In this case the random Fourier feature
approximation is made as
$\vec{\phi}_\mathrm{NS} = \vec{\phi}_1 + \vec{\phi}_2$ where the frequencies used
in $\vec{\phi}_1$ and $\vec{\phi}_2$ are drawn from a joint distribution,
$\pi_K(\omega_1,\omega_2)$.  In this case we take an analogue of the square
exponential kernel with
\begin{equation*}
\pi_K(\omega_1,\omega_2)=\mathcal{N}(0,b_1)[\omega_1]\times \mathcal{N}(0,b_2)[\omega_2].
\end{equation*}
Our GRB signal model is thus defined explicitly as
\begin{widetext}
\begin{align}
  \label{eq:1}
  \log S(t | b_1,b_2, \sigma_1, \sigma_2, \vec{\omega}_1,\vec{\omega}_2, \vec{\beta}_1, \vec{\beta}_2 ) = \left[\sum_{i=1}^2 \frac{\sigma_i}{\sqrt{k}}\cos(b_i \vec{\omega}_i \otimes t) \right] \cdot \vec{\beta}_1 + \left[\sum_{i=1}^2 \frac{\sigma_i}{\sqrt{k}}\sin(b_i \vec{\omega}_i \otimes t) \right]\cdot \vec{\beta}_2
\end{align}
\end{widetext}
\subsection{Background signal model}
To complete our forward model we need finally to specify a
distribution for the expected background signals to be received by
each detector.  For simplicity we suppose here that each background is
a constant function of Poisson counts with positive intensity, i.e.,
$B_i(t) = B_i$.  Again, more complicated background signals can
readily be handled within this hierarchical framework.

\subsection{Directed acyclic graph view}
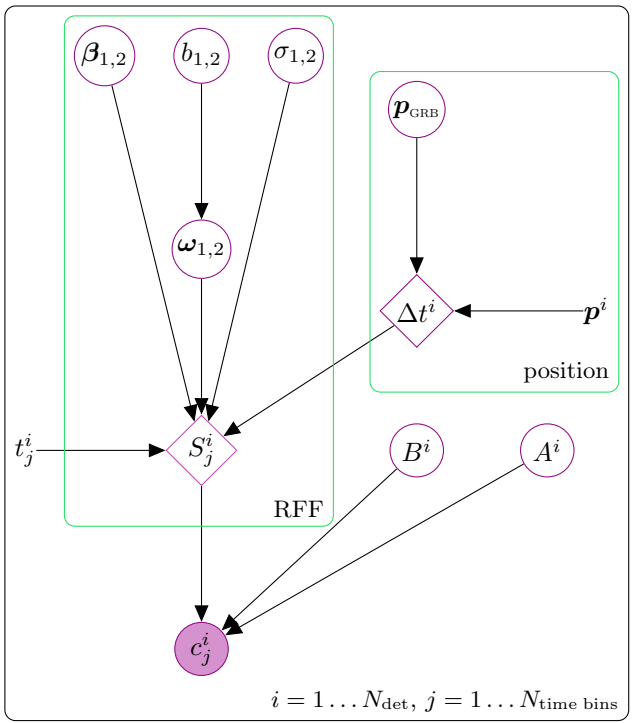
\begin{figure}
    \centering
\begin{tikzpicture}[x=1.7cm,y=1.8cm]


  \node[obs][fill=light,draw=dark]                   (C)      {$c_j^i$} ; %
  \node[det, above=of C][draw=mid]    (S)      {$S_j^i$}; %
  \node[const, left=of S]    (t)  {$t_j^i$}; %
  \node[latent, right=2cm of S][draw=dark]     (bkg)      {$B^i$} ; %
  \node[latent, right=1cm of bkg][draw=dark]     (amp)      {$A^i$} ; %
  \node[det, above=1cm of bkg][draw=dark]     (dt)      {$\Delta t^i$} ; %
  \node[const, right=of dt]    (p)  {$\vec{p}^i$}; %
  \node[latent, above=of S][draw=dark]     (w)  {$\vec{\omega}_{1,2}$}; %
  \node[latent, above=of w][draw=dark]     (b)  {$b_{1,2}$}; %
  \node[latent, right=.5cm of b][draw=dark]     (sig)  {$\sigma_{1,2}$}; %
  \node[latent, left=.5cm of b][draw=dark]     (beta)  {$\vec{\beta}_{1,2}$}; %
  \node[latent, above=of dt][draw=dark]     (grb)  {$\pgrb$}; %

\edge {beta, w, sig} {S};
\edge {grb} {dt};
\edge {b} {w};

\edge {bkg} {C};

\edge {S} {C};

\edge {t} {S};

\edge {p} {dt};

\edge {dt} {S};
\edge {amp} {C};

\plate [color=green] {plate1} {%
  (beta) (b) (S) (w) (sig)%
} {RFF};
\plate [color=green] {plate2} {%
  (p) (dt) (grb)%
} {position}

\plate [color=black] {plate3} {%
  (plate1) (plate2)%
  (amp) (bkg) (C) (t)%
  } { $i=1\dots N_{\mathrm{det}}$, 
  $j=1 \dots N_{\mathrm{time\;bins}}$}

\end{tikzpicture}
\caption{The directed acyclic graph of our model. }
\label{fig:dag}
\end{figure}

In Figure \ref{fig:dag} below we illustrate our hierarchical model as
a directed acyclic graph. This diagrammatic mode of presentation highlights the connections
between the parameters at each level of the hierarchical model and the data.  Here, open circles
represent latent variables to be inferred while closed circles indicate
data. Quantities not enclosed in circles are constant information that
do not exhibit measurement uncertainty. Finally, quantities enclosed
in diamonds represent deterministic combinations
of latent variables and/or fixed quantities.

\section{Validation}
In order to demonstrate and validate our method, we have created a
simulation
package\footnote{\faGithub~\href{https://github.com/grburgess/pyipn}{https://github.com/grburgess/pyipn}}
that allows for the virtual placement of GRB detectors throughout the
solar system and creates energy-independent Poisson events sampled
from a given light curve model. Using this package, we select a number
of representative combinations of satellite configurations and light
curve shapes and perform analyses on mock data simulated under each
setting. We additionally compute the localization of the simulated GRB
with the cIPN method to serve as a comparison. Our procedure for
simulating mock datasets is described in Appendix \ref{sec:sims}. In
the following we first demonstrate the ability of the {\tt nazgul} model to
fit a single time-series data set, and then proceed to fit two and
three detector configurations to recover time delays and perform
localizations.  By repeating the simulation and localization process
many times over, we confirm that the associated uncertainty intervals
of the {\tt nazgul} algorithm are well calibrated in a Frequentist (long-run)
sense, i.e., that each quoted uncertainty interval contains the true
source location with a frequency close to its nominal level.

\subsection{Validation of pulse fitting with RFFs}
\begin{figure}[ht]
  \centering
  \includegraphics[]{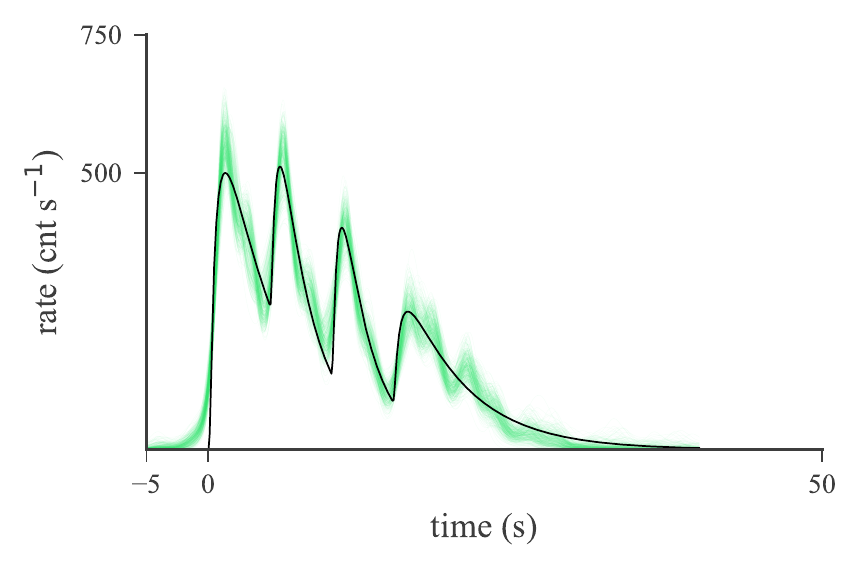}
  \caption{The latent (black) and inferred (green traces) source light
    curve.}
  \label{fig:simonefit}
\end{figure}

\begin{figure}[ht]
  \centering
  \includegraphics[]{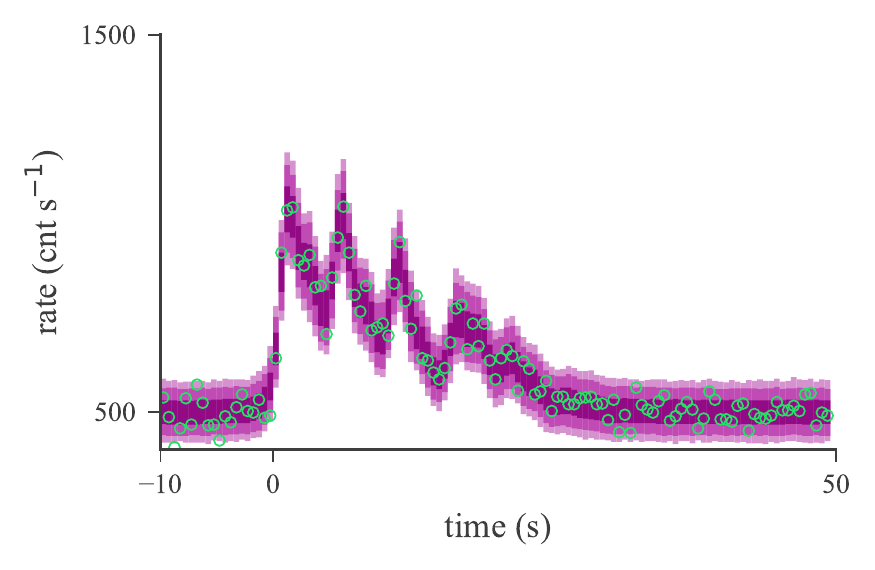}
  \caption{The count rate data and PPCs from the single detector
    fit. The 1, 2 and 3$\sigma$ regions are plotted in purple along with
    the data shown in green.  }
  \label{fig:simoneppc}
\end{figure}

Here we demonstrate the ability of RFFs to model simulated light
curves with Poisson count noise. A GRB is simulated with four
overlapping pulses each beginning five seconds after the pulse
preceding it. In Figure \ref{fig:simonefit} we can see the posterior
traces of the fit compared with the true light curve. We can check the
reconstruction of the light curves via posterior predictive checks
\citep[][PPCs]{Gabry:2019aa} where counts in each time bins are
sampled from the inferred rate integrated over the posterior (see
Figure \ref{fig:simoneppc}).

We have checked that we are able to recover a variety of pulse shapes
and combinations. Thus, even without the aid of a secondary signal
from another detector, RFFs can learn the shape of the latent signal. 

When the strength of the signal is decreased, we notice that the
inferences become less constraining, but the overall ability of the
model to reproduce the latent signal is sufficient to find properties
of the light curve that can be time-delayed.

\subsection{Two detector validation}
The simulation for only two detectors allows us to validate the
simplest configuration of detectors possible which results in a single
annulus on the sky. For this simulation, the configuration consists of
one detector in low-Earth orbit and another at an altitude mimicking
an orbit in L1. The light curve shape chosen consists of two pulses
and the detectors are given effective areas that differ by a factor of
two. Each light curve is sampled at a different cadence (100 ms and
200 ms). Details are given in Tables \ref{tab:sim1}-\ref{tab:grb1}.

\begin{table*}[ht]
  \centering
  \caption{Detector simulation parameters. See Appendix \ref{sec:sims} for details.}
  \vspace{-0.3cm}
\begin{tabular}{cccccc}
  \hline
   name &  $\Delta t$ (s) &      altitude (km) &   position (ra, dec) &   pointing (ra,dec)&  effective area (arb. units)\\
  \hline
  \hline
   d$_{1}$ &    0 &  1500000 &   40, 5  &  20, 40 &  1 \\
    d$_{2}$ &    3.87 &      500 &  60, 10 &  20, 40 &  0.5 \\
  \hline
\end{tabular}
  \label{tab:sim1}
\end{table*}

\begin{table}[ht]
  \centering
  \caption{GRB simulation parameters. See Appendix \ref{sec:sims} for details.}
  \vspace{-0.3cm}
  \begin{tabular}{l|l}
  \hline
    location (ra, dec) &    20.0,40.0 \\
  \hline\hline
    $K$ (ph s$^{-1}$)  & 400, 400 \\
    $\tau_{\mathrm{s}}$ (s) & 0, 4 \\
    $\tau_{\mathrm{r}}$ (s) & 0.5,0.5 \\
    $\tau_{\mathrm{d}}$ (s) &  4,3 \\
   \hline    
\end{tabular}

  \label{tab:grb1}
\end{table}

We have found that choosing $k=25$ features is sufficient for our
purposes. Specific details of the fitting algorithm can be found in
Appendix \ref{sec:fitting}. For the two detector fit, we compute
credible region contours by computing the credible regions of
$\Delta t$ and translating these to annuli on the sky. We note that this is
not the same as will be done for fits with more detectors.

\begin{figure}[ht]
  \centering
  \includegraphics[]{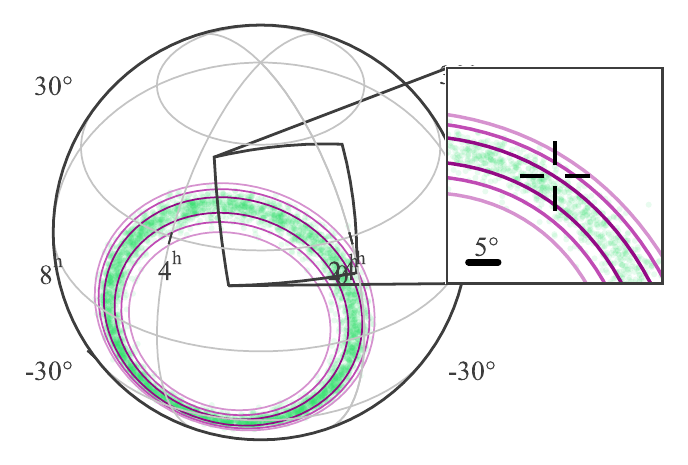}
  \caption{Skymap of the two detector validation. The purple contours
    indicate the 1, 2, and 3$\sigma$ credible regions with increasing
    lightness. The posterior samples are shown in green and the
    simulated position is indicated by the cross-hairs.}
  \label{fig:2df}
\end{figure}

\begin{figure}[ht]
  \centering
  \includegraphics[]{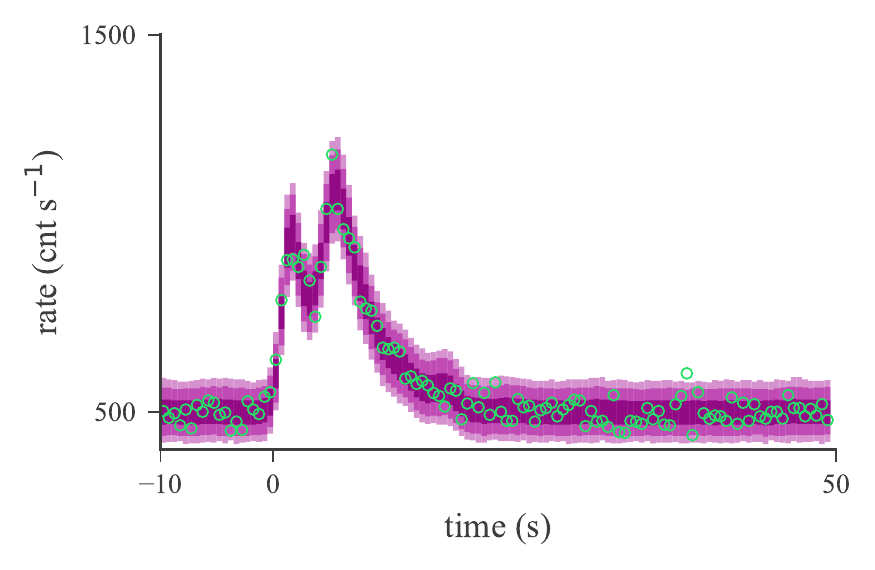}
  \caption{Posterior predictive checks from fit in one of the
    detectors. The 1, 2 and 3$\sigma$ regions are plotted in purple along
    with the data shown in green.}
  \label{fig:ppc1}
\end{figure}

\begin{figure*}[h]
  \centering
  \includegraphics[]{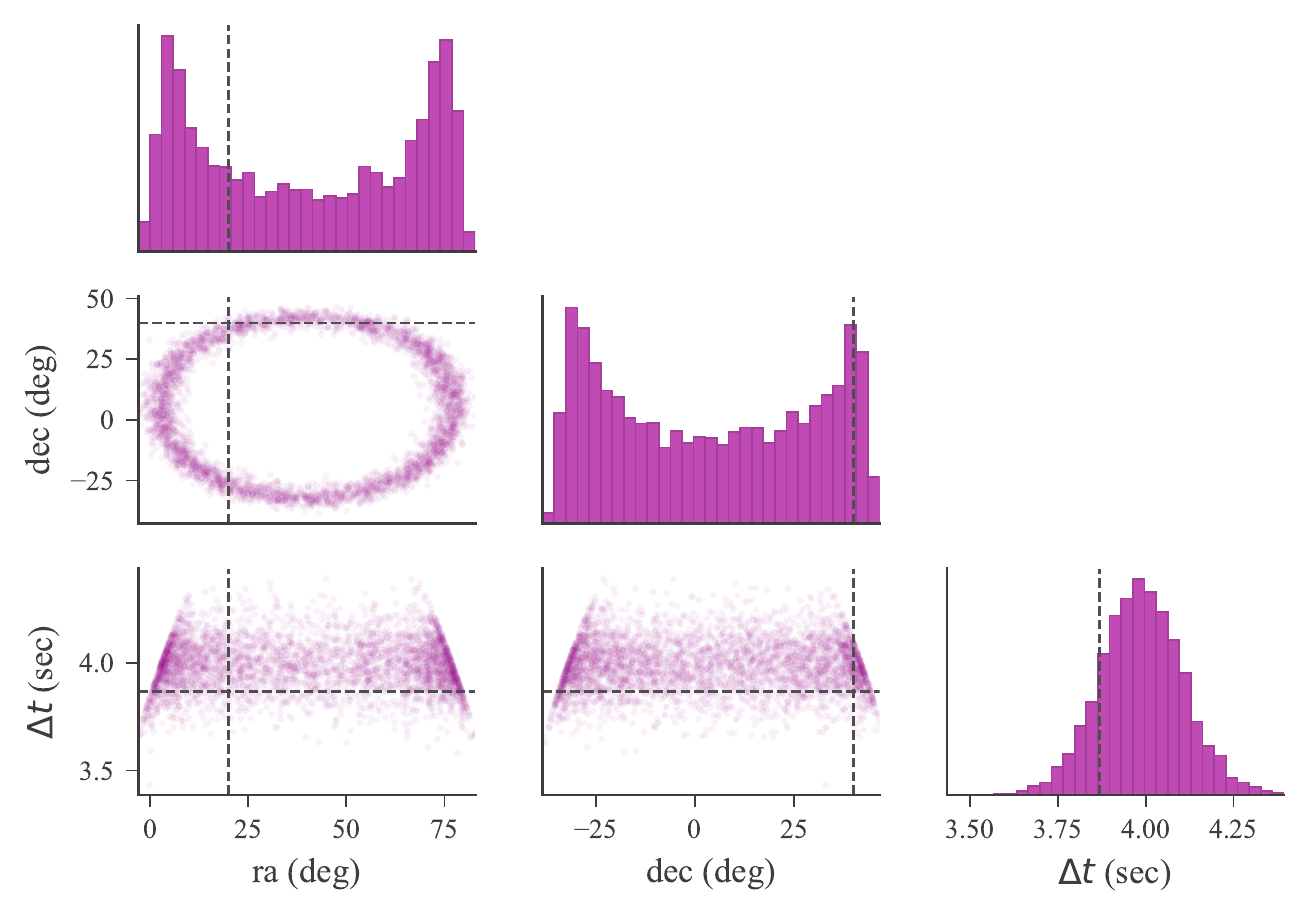}
  \caption{Posterior density pair plots demonstrating the correlations
    between position an time delay. Simulated values are shown via
    grey lines.}
  \label{fig:hist1}
\end{figure*}

As can be seen in Figure \ref{fig:2df}, the algorithm correctly finds
annuli enclosing the true position of the simulated GRB. The latent
signal is recovered as shown in Figure \ref{fig:2df} and the PPCs
adequately reproduce the data as seen in Figure \ref{fig:ppc1}. We can
examine the correlations between position and time that occur
naturally due to our forward model approach in Figure
\ref{fig:hist1}. By using different combinations of temporal binnings,
we find that the result is not dependent on this parameter unless the
binning is too large to resolve the time-delay. Even in this case;
however, we find that the result is only marginally affected with
larger uncertainties on the time-delay.

We have repeated the experiment with same setup but different Poisson
realizations of the light curves 500 times to ensure that our contours
are statistically robust, i.e, that a give credible region is a proper
representation of the probability that the latent value is
recovered. Figure \ref{fig:joy} shows the marginals of $\Delta t$ for each
fit to these simulations and clearly demonstrates that different
Poisson realizations of the same light curve can lead to widely
different $\Delta t$. Nevertheless, we can compute the number of times the
simulated $\Delta t$ falls within a given credible region with the
expectation that the true value should fall $M$ times within CR$_{p}$
such that $p\simeq M/N_{\mathrm{sims}}$. We demonstrate that this condition
is held in Figure \ref{fig:cred1}.

By simulating a grid of pulse intensities spanning three orders of
magnitude, we can investigate the {\tt nazgul} algorithm's ability to
reconstruct the time-delay for a variety of pulse intensities. The
same geometry and pulse shape is used except that the intensities of
the pulses are modified to yield source significances ranging from
$ \sim 1\sigma $ to $\sim 200 \sigma$ Here, we define significances with the
likelihood ratio of \citet{Li:1983aa} and compute the significance
over the peak flux duration of the pulse (see Appendix
\ref{sec:sig}). Fits are performed by binning each light curve to 100
ms. At low significance, the {\tt nazgul} method is very uncertain
about the true value of $\Delta t$, but at $\sim 5\sigma$ the method begins to have
a trend at not only being accurate at recovering the true value but
also becomes increasingly precise (see Figure
\ref{fig:significance}). Importantly, the {\tt nazgul} method and
resolve $\Delta t$ \emph{appreciably below} the resolution of the data for
brighter simulated GRBs. These results are difficult to generalize as
pulse shape and detector geometry can also play a role.

\begin{figure}[h]
  \centering
  \includegraphics[]{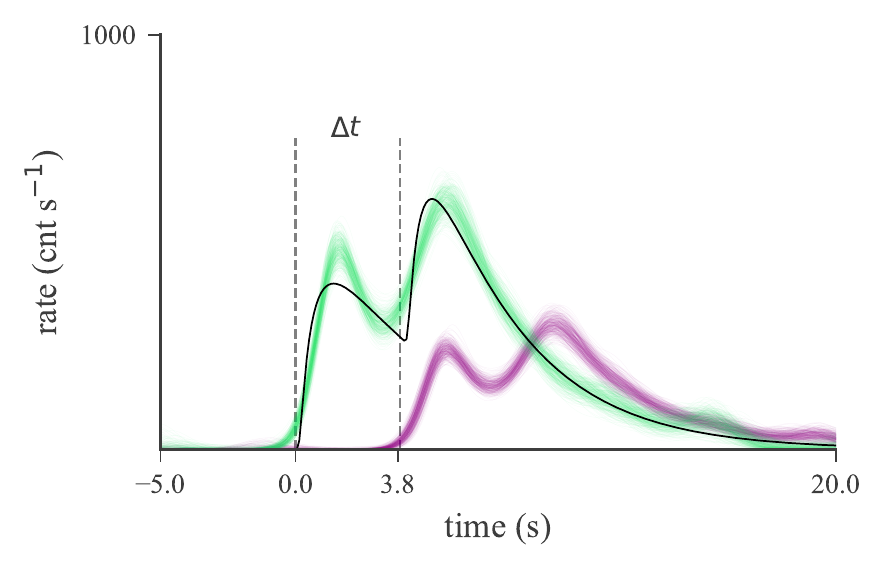}
  \caption{The inferred posterior traces of the pulse models to in the
    two detectors (green and purple) compared with the non-delayed
    simulated pulse model.}
  \label{fig:lc2}
\end{figure}

\begin{figure}[ht]
  \centering
  \includegraphics[]{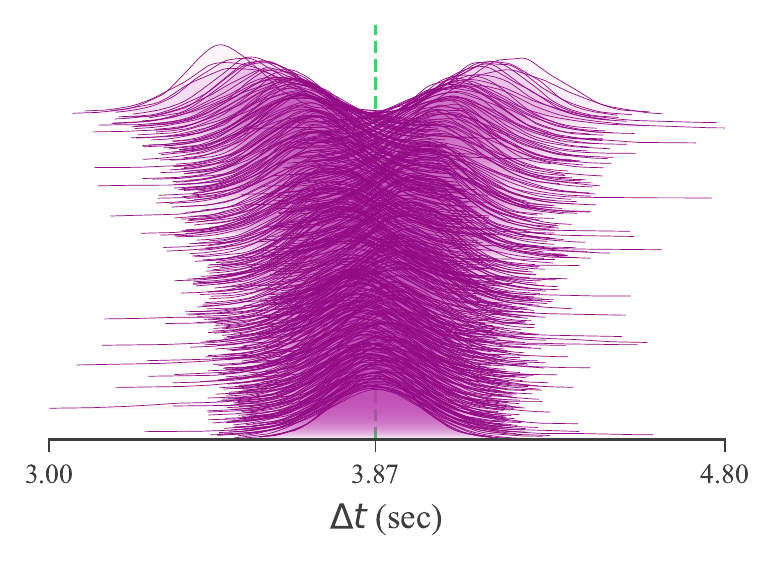}
  \caption{The marginal distributions (purple) of fits to each two
    detector simulation ordered by their median's distance to the true
    value (green)}
  \label{fig:joy}
\end{figure}

\begin{figure}[ht]
  \centering
  \includegraphics[]{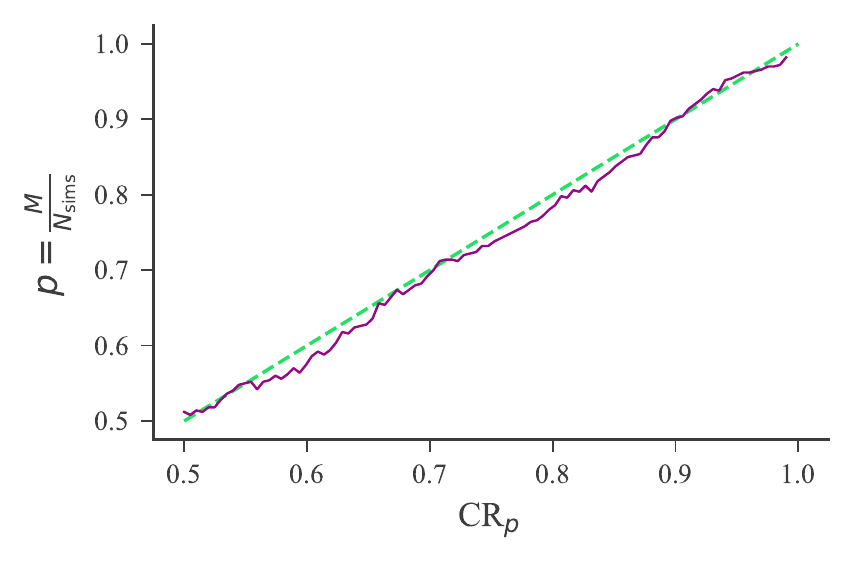}
  \caption{Credible regions of $\Delta t$ versus the fraction of times the
    true, simulated value is within that credible region. The green
    line represents a perfect, one-to-one relation.}
  \label{fig:cred1}
\end{figure}

\begin{figure}[ht]
  \centering
  \begin{subfigure}[]{\includegraphics[]{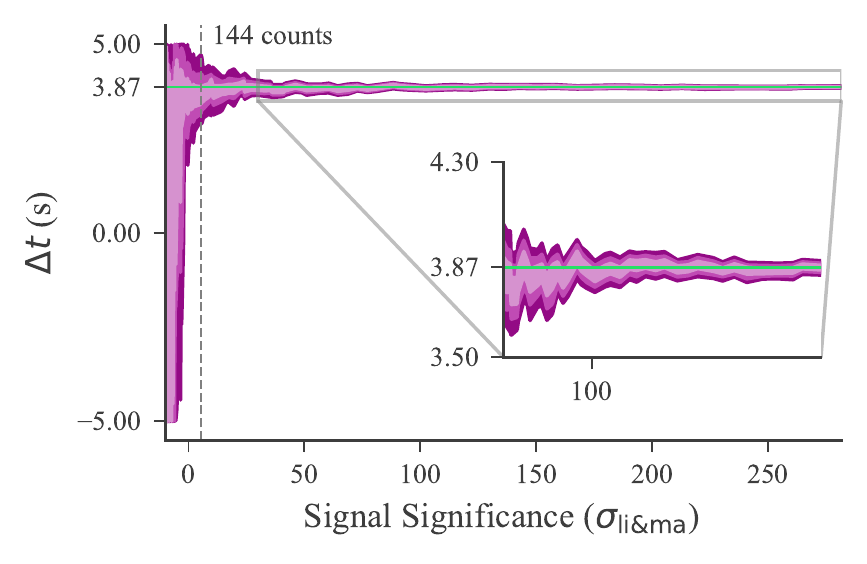}}
  \end{subfigure}
    \begin{subfigure}[]{\includegraphics[]{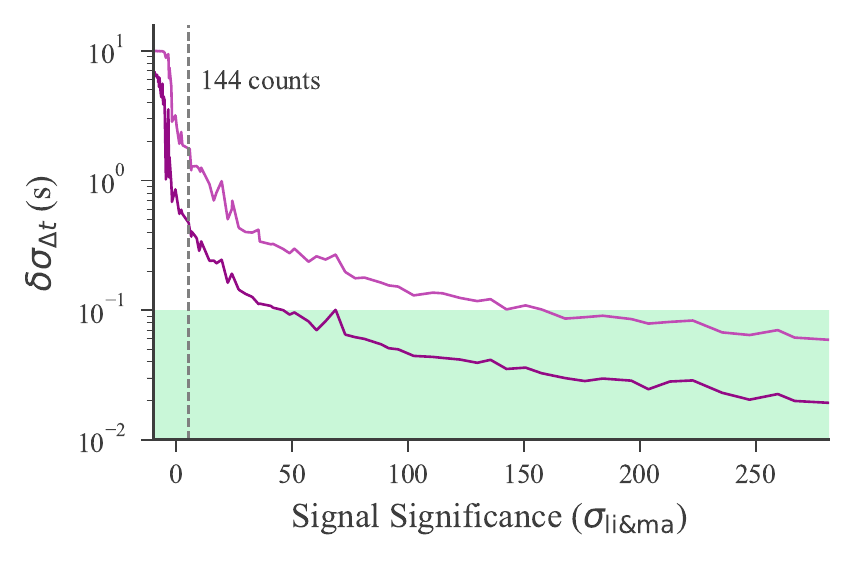}}
    \end{subfigure}
    \caption{ (a) The 1, 2, and 3$\sigma$ uncertainties (purple) of
      $\Delta t$ as a function of significance compared with the simulated
      value (green). (b) The width of the 1 and 2 $\sigma$ uncertainties
      for the {\tt nazgul} and cIPN (purple and green respectively) as
      a function of significance. The green region is below the
      resolution of the data and the dashed line appears at
      $\sim 5 \sigma$ where the number of background-subtracted counts in the
      peak are displayed.}
  \label{fig:significance}
\end{figure}

\subsubsection{Comparison with the classical method}
The cIPN method of cross-correlating two background-subtracted light
curves has had great success in localizing GRBs and other $\gamma$-ray
transients \citep[e.g.][]{Cline:1982aa, Hurley:1998aa}. However, part
of the motivation of our work is to provide a newer algorithm with
statistically robust uncertainties. Thus, we now run the classical
algorithm on our simulated data sets and compare the results. It must
be noted that this algorithm relies on two procedures which have known
issues, namely, background subtraction and a Gaussian approximation to
the Poisson distribution \citep[for details see][Appendix
\ref{sec:ccr}]{Palshin:2013aa,Hurley:2013aa}.

\begin{figure}[ht]
  \centering
  \begin{subfigure}[]{\includegraphics[]{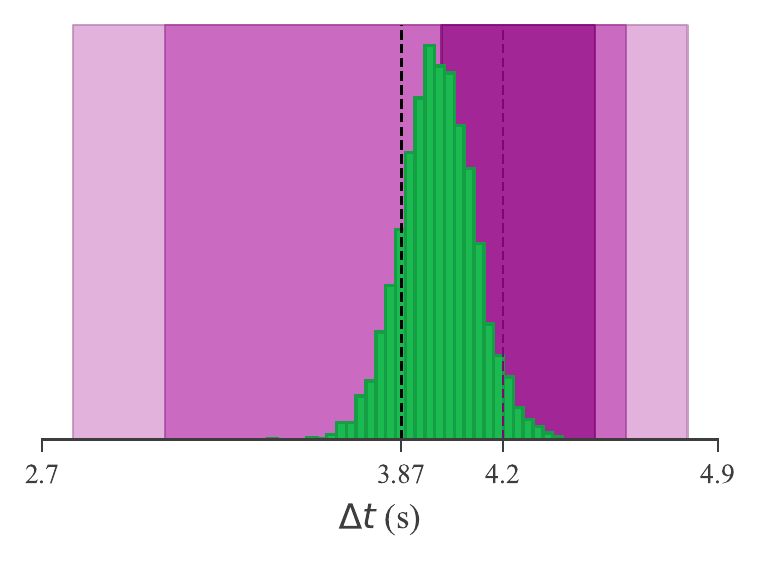}}
  \end{subfigure}
  \begin{subfigure}[]{\includegraphics[]{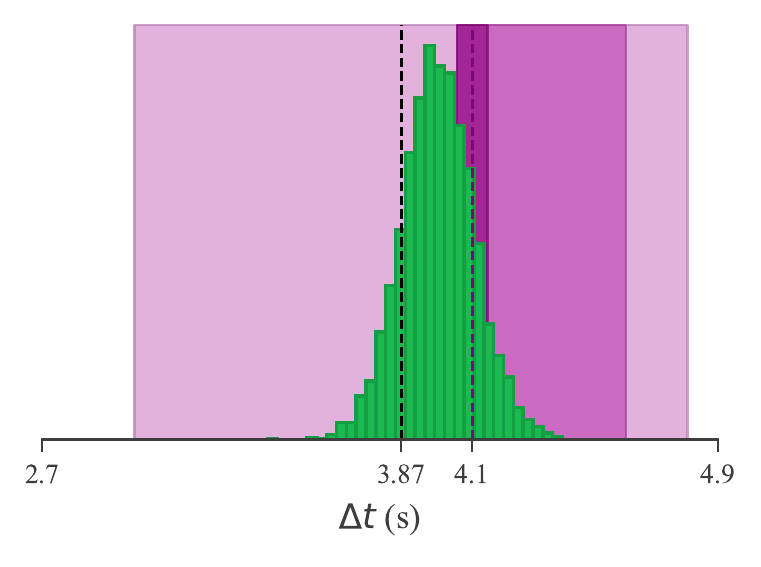}}
    \end{subfigure}
    \caption{The marginal distribution of $\Delta t$ from the {\tt nazgul} fit
      (green) compared with the 1, 2 and 3$\sigma$ confidence intervals
      (purple) from the cIPN method for two different resolutions. The
      coarse resolution used for the {\tt nazgul} fit (a) and a finer 50 ms
      resolution (b). The true value is shown as a black line and
      the mean value from the cIPN as a purple line. }
  \label{fig:singcomp}
\end{figure}

First we examine the method on our example simulation.  While our new
algorithm is independent of binning and pre/post- source temporal
selection, we find that the same is not true the cIPN
cross-correlation approach, and different selections can have vastly
different results. Thus, we choose a selection, having knowledge of
the simulation parameters\footnote{This is obviously not the case for
  real data and we expect our findings to not probe the systematics of
  source selection which may be considerable.} that encompasses the
burst interval and has enough lag between the signals to fully map out
the so-called reduced $\chi^2$-contour as a function of lag. We also
chose different combinations temporal binnings for the light curves
ranging from 50-200 ms; the coarser corresponding to what was used for
the {\tt nazgul} method. The results from two of the fits at different
resolution is show in Figure \ref{fig:singcomp}. We can see that the
choice of binning immensely influences the confidence
intervals. Additionally, the 1$\sigma$ region misses the true value while
the 3$\sigma$ region is much larger than the {\tt nazgul} posterior.

\begin{figure}[ht]
  \centering
  \includegraphics[]{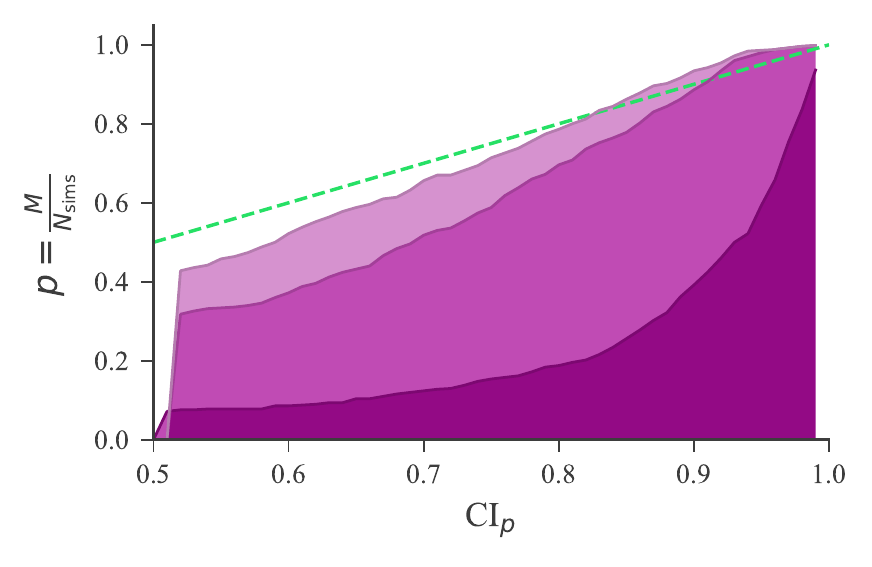}
  \caption{Confidence intervals of $\Delta t$ versus the fraction of times the
    true, simulated value is within that credible intervals. The green
    line represents a perfect, one-to-one relation. Here, coarser
    binnings are represented by lighter purple colors.}
  \label{fig:cred2}
\end{figure}

To see if these results can be generalized, we perform
cross-correlation on each of the 500 simulations. We can repeat the
exercise above and compute the fraction of fits that capture the
simulated value given a confidence interval\footnote{We note that
  there is a marked difference between a credible region and credible
  interval but for our purpose to calibrate coverage and with the
  Gaussian marginals we obtain, the difference is small.}. Figure
\ref{fig:cred2} shows that for different temporal binnings, the
credible intervals do not exhibit appropriate coverage. For coarser
binnings, the coverage begins to approach expectations. This is likely
due to the number of counts in each interval increasing for these
binnings and thus satisfying the Gaussian approximation of the Poisson
distribution. This poses severe difficulties for detector geometries
which would need high-resolution to find sub-second time delays
between light curves. We note that there are some combinations of
resolution that result in adequate coverage in the $3\sigma$ intervals, but
in situations involving real data, it would be impossible to know
which combinations would be required.

\begin{figure}[ht]
  \centering
  \includegraphics[]{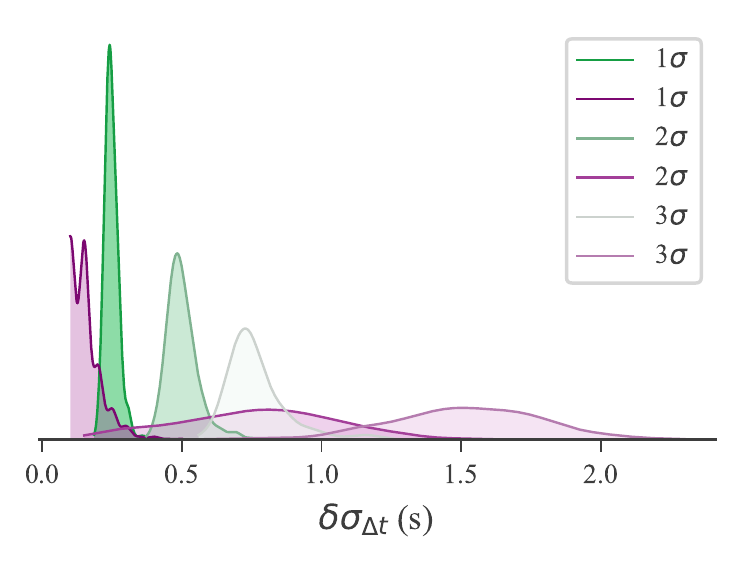}
  \caption{Distributions of the relative widths of the uncertainties
    on $\Delta t$ at three different credible levels from the classical
    (purple) and new (green) algorithm fits to simulations. }
  \label{fig:width}
\end{figure}

Moreover, we can examine the relative widths of the uncertainties of
the two methods at various confidence intervals. We find that while
our algorithm is well behaved (see Figure \ref{fig:width}), the
1$\sigma$ uncertainties of the cIPN method are smaller and (as shown above)
underestimated, while the $3\sigma$ uncertainties are appreciably larger
and likely over-estimated.

\begin{figure}[ht]
  \centering
  \begin{subfigure}[]{\includegraphics[]{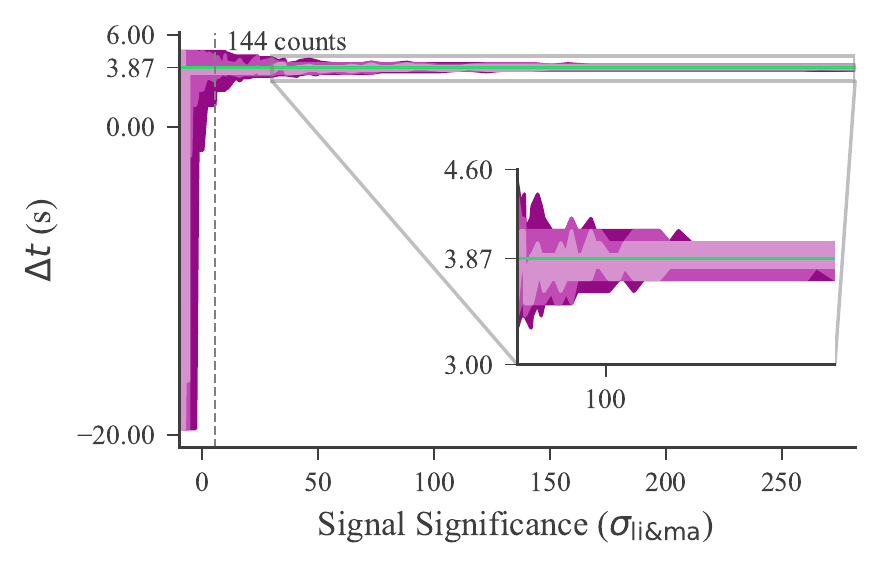}}
  \end{subfigure}
    \begin{subfigure}[]{\includegraphics[]{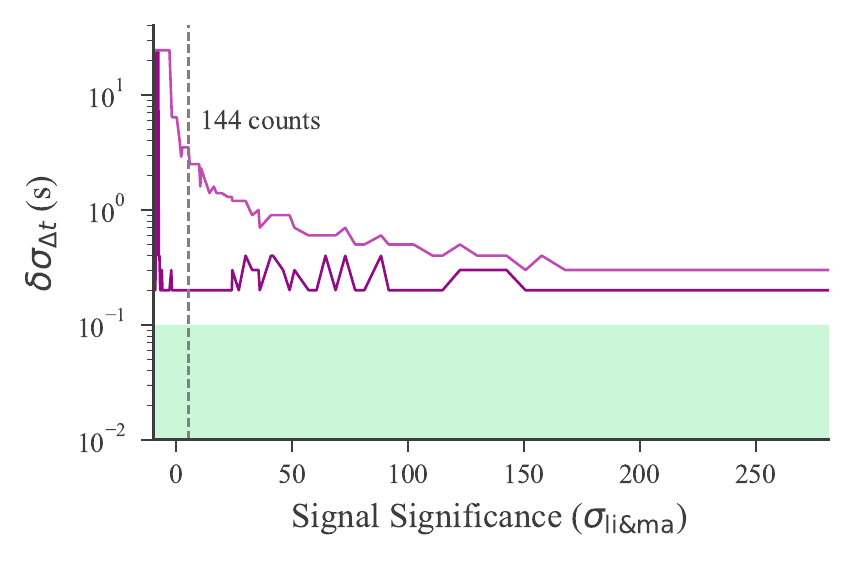}}
    \end{subfigure}
    \caption{ (a) The 1, 2, and 3$\sigma$ uncertainties (purple) of
      $\Delta t$ as a function of significance compared with the simulated
      value (green). (b) The width of the 1 and 2 $\sigma$ uncertainties
      for the cIPN (purple) as a function of significance. The green
      region is below the resolution of the data and the dashed line
      appears at $\sim 5 \sigma$ where the number of background-subtracted
      counts in the peak are displayed.}
  \label{fig:significance_cc}
\end{figure}

We also examine how the cIPN method behaves with burst intensity. In
Figure \ref{fig:significance_cc} we compare the performance against
the {\tt nazgul}. All light curves were binned to 100 ms resolution. We again
note, that in the actual use of the cIPN method, temporal resolution
is chosen by hand to account for weak signals, but our procedure is
chosen so that we can have direct comparisons. The $1 \sigma$ behavior
typically has a width of twice the resolution as expected, but even in
the low-count regime (with some exceptions). The $2 \sigma$ regions are
better behaved similar to the {\tt nazgul} method, but are much wider.

\begin{figure}[ht]
  \centering
  \includegraphics[]{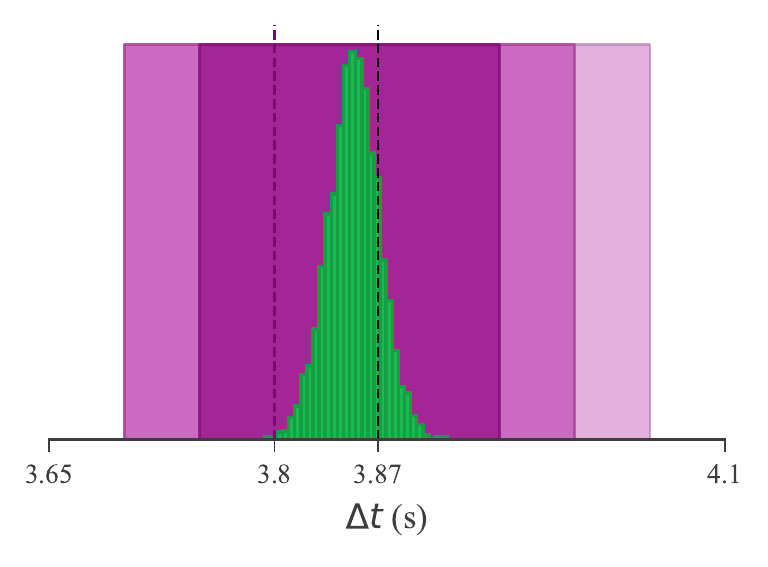}
  \caption{The marginal distribution of $\Delta t$ from the {\tt nazgul} fit
    (green) compared with the 1, 2 and 3$\sigma$ confidence intervals
    (purple) from the cIPN method for two different resolutions. The
    coarse resolution used for the {\tt nazgul} fit. The true value is shown
    as a black line and the mean value from the cIPN as a purple
    line.}
  \label{fig:dtbright}
\end{figure}

Noting that the cIPN algorithm relies on a central limit theorem
asymptotic of having many counts in a temporal interval, we simulate
the same geometry but increase the flux by a factor of ten. We then
repeat both methods on this brighter GRB. First, we fit with the {\tt
  nazgul} algorithm with resolutions of 100 ms and 200 ms for each
light curve respectively. We then use the cIPN method with a
resolution of 50ms. Figure \ref{fig:dtbright} again compares the
uncertainties of these two fits. In this case, the cIPN uncertainties
behave properly, i.e., they are accurate, but surprisingly, the {\tt
  nazgul} uncertainties are much more precise. Increasing the
resolution to 50 ms for the {\tt nazgul} has no effect and the
precision of the result, but decreasing the resolution of the cIPN
method to that used for the initial {\tt nazgul} result increases the
uncertainties by a fact of $~2$ as expected. However, it is possible
to obtain sub-resolution uncertainties with the cIPN method by fitting
a Gaussian to the reduced-$\chi^2$ curves as can be seen in Figure 3 of
\citep{Hurley:1999aa}.

We again examine the this behavior on multiple simulations. The {\tt nazgul}
maintains its adequate coverage, but the cIPN now begins to exhibit
coverage that is semi-proper with fine temporal binnings. Though, the
cIPN still underestimate $1\sigma$ intervals and overestimate
$3\sigma$ intervals. The intervals for coarser binnings are too large in
all cases. More importantly, the distribution of the widths for the
cIPN method are appreciably larger than those of the {\tt nazgul}. While we do
not intend to do a comprehensive calibration of the cIPN method, we
can demonstrate that there are issues with its ability to provide
reliable uncertainties even in our toy setup. However, our newer
method provides robust uncertainties with more precise $3\sigma$
contours. Also, it should be noted that the combinations of temporal
resolutions and GRB intensities chosen do not always reflect the
actual choices which are made by the IPN team. For example, if a weak
GRB is being analyzed, coarser resolutions are chosen in an attempt to
satisfy the number of counts required to satisfy the asymptotic
conditions of the central limit theorem. Nevertheless, we find it
important to demonstrate the frailty of these asymptotic methods as
they are utilized in many places in the low-count regime of gamma-ray
astrophysics.

\begin{figure}[ht]
  \centering
  \includegraphics[]{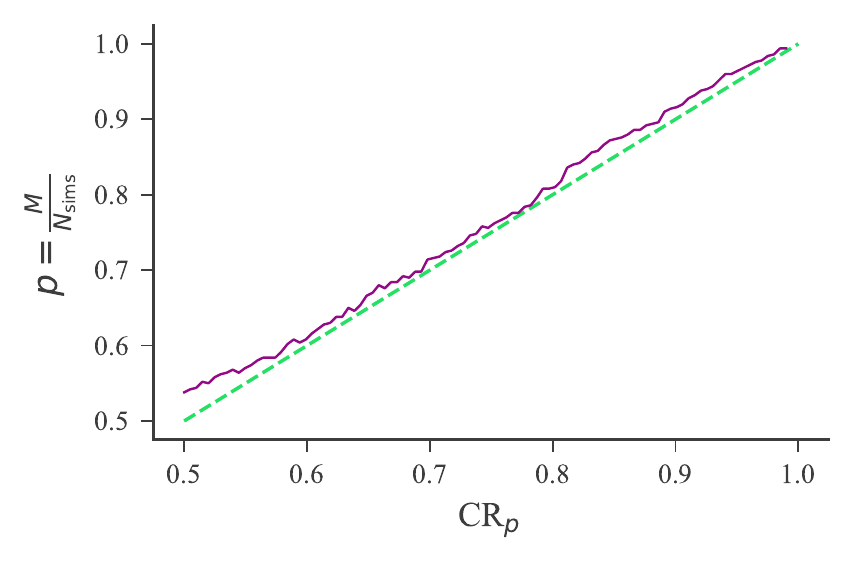}
  \caption{Credible regions of $\Delta t$ versus the fraction of times the
    true, simulated value is within that credible region. The green
    line represents a perfect, one-to-one relation.}
  \label{fig:covbright}
\end{figure}

\begin{figure}[ht]
  \centering
  \includegraphics[]{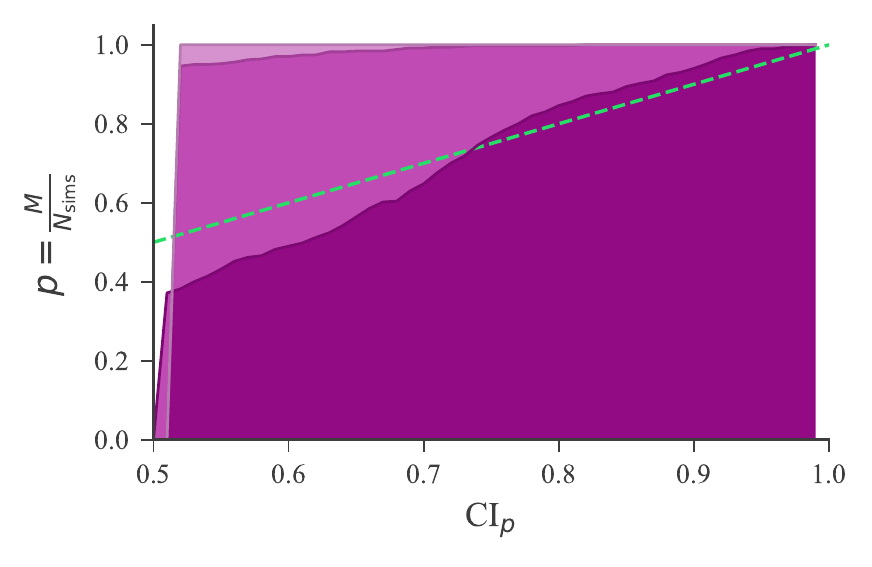}
  \caption{Confidence intervals of $\Delta t$ versus the fraction of times the
    true, simulated value is within that credible intervals. The green
    line represents a perfect, one-to-one relation. Here, coarser
    binnings are represented by lighter purple colors.}
  \label{fig:covccbright}
\end{figure}

\begin{figure}[ht]
  \centering
  \includegraphics[]{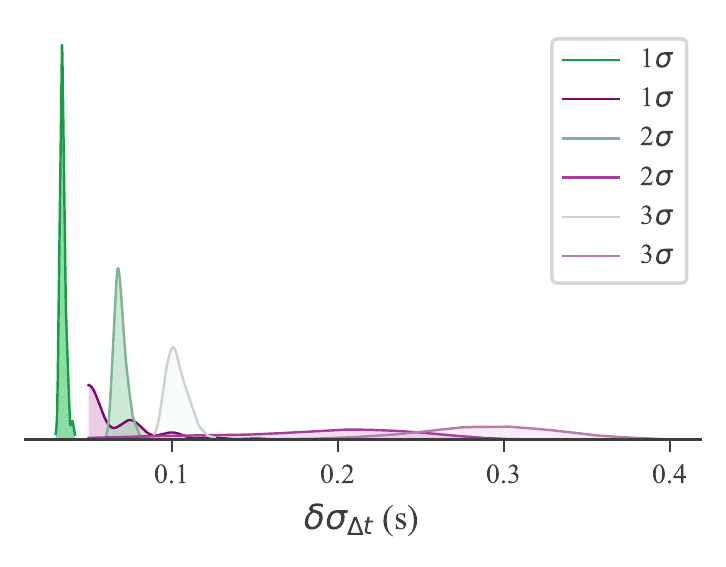}
  \caption{Distributions of the relative widths of the uncertainties
    on $\Delta t$ at three different credible levels from the classical
    (purple) and new (green) algorithm fits to simulations.}
  \label{fig:widthbright}
\end{figure}

\subsection{Three detector validation}
Extending upon our two detector validation, we now add a third
detector to break the single-annulus degeneracy and further verify our
algorithm. The above verification process is repeated with a new
detector geometry as shown in Tables
\ref{tab:sim2}-\ref{tab:grb2}. This time, a single pulse is simulated
again with all three detectors having different effective areas which
are fitted for.

\begin{table*}[ht]
  \centering
  \caption{Detector simulation parameters. See Appendix \ref{sec:sims} for details.}
  \vspace{-0.3cm}
  \begin{tabular}{cccccc}
  \hline
   name &  $ \Delta t$ (s) &  altitude (km) & position (ra, dec)& pointing (ra, dec) & effective area \\
    \hline
    \hline
d$_1$ &    0 &  1500000 & 40, 5 & 100, -30 &    1 \\
d$_2$ &    1.93 &  500  & 60, 10 &   100, -30 &  0.75 \\
d$_3$ &    1.68 & 153000 & 160, -10 &  100, -30 &   1.50 \\
 \hline
  \end{tabular}
  \label{tab:sim2}
\end{table*}

\begin{table}[ht]
  \centering
  \caption{GRB simulation parameters. See Appendix \ref{sec:sims} for details.}
  \vspace{-0.3cm}
  \begin{tabular}{l|l}
  \hline
    location (ra, dec) &  100, -30 \\
    \hline\hline
    $K$  ph s$^{-1}$ &  500 \\
    $\tau_{\mathrm{r}}$ (s) &     1 \\
    $\tau_{\mathrm{d}}$ (s) &     7 \\
    \hline
  \end{tabular}
  \label{tab:grb2}
\end{table}

\begin{figure}[ht]
  \centering
  \includegraphics[]{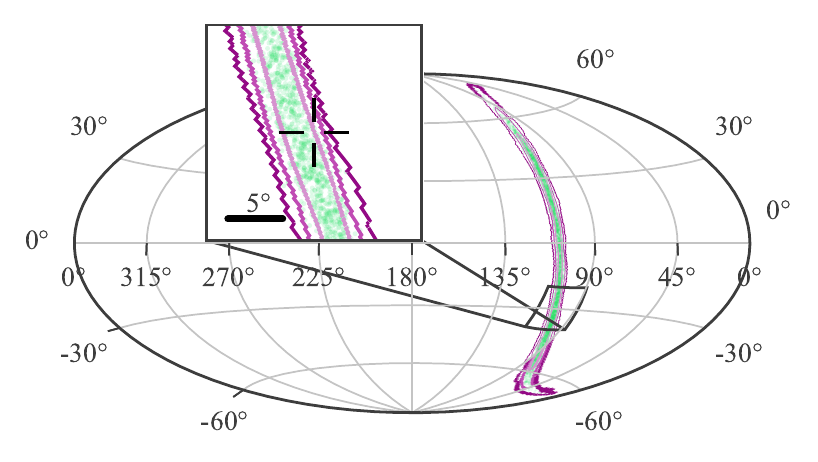}
  \caption{Skymap of the three detector validation. The purple
    contours indicate the 1, 2, and 3$\sigma$ credible regions with
    increasing lightness. The posterior samples are shown in green and
    the simulated position is indicated by the cross-hairs.  }
  \label{fig:3df}
\end{figure}

The {\tt nazgul} algorithm clearly infers the correct position as shown in
Figure \ref{fig:3df}. However, the uncertainty regions do not mimic
the diamond shaped uncertainties created by intersecting
two annular uncertainty regions as is done by the cIPN
\citep{Sinha:2002aa}. This is expected as we have demonstrated above
that the two detector uncertainties regions are not correctly
estimated and the way that these intersections are treated in the cIPN
is reduced to a geometric rather than a statistical problem. The fact
that our likelihood naturally includes all detector baselines in a
single fit and treats the position of the GRB as a parameter rather
than a downstream calculated quantity allows us to properly compute
these uncertainty regions without reliance on asymptotic uncertainty
propagation.

\begin{figure}[ht]
  \centering
  \includegraphics[]{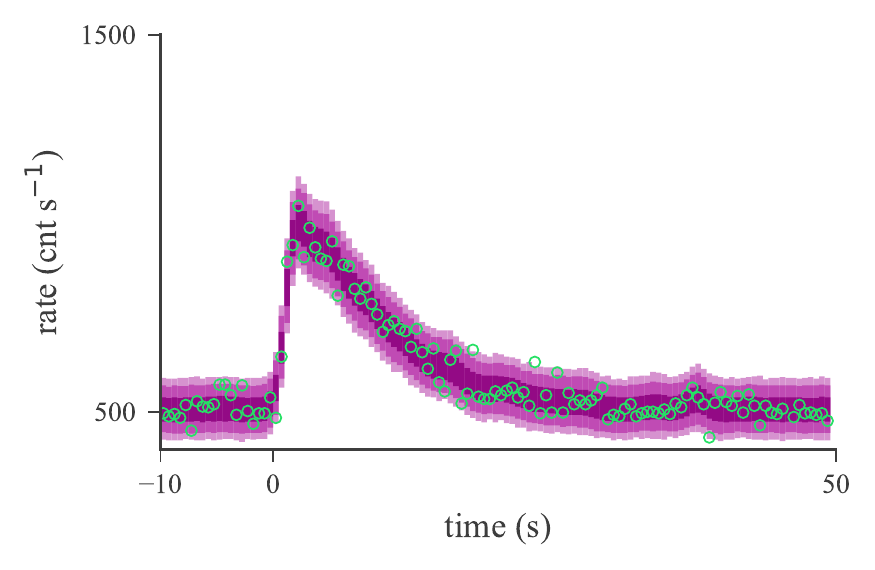}
  \caption{Posterior predictive checks from fit in one of the
    detectors. The 1, 2 and 3$\sigma$ regions are plotted in purple along
    with the data shown in green. }
  \label{fig:ppc2}
\end{figure}

\begin{figure*}[h]
  \centering
  \includegraphics[]{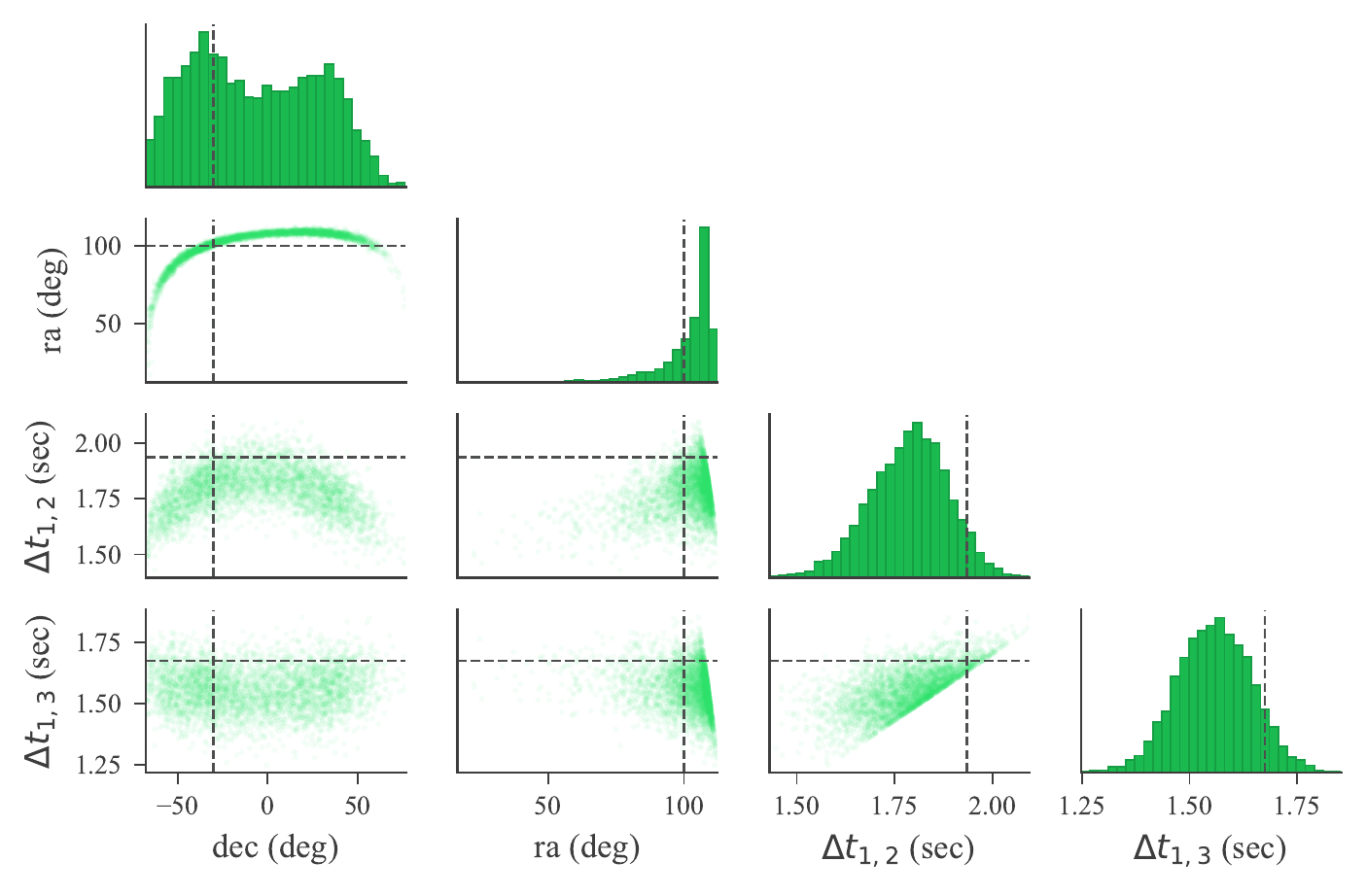}
  \caption{Posterior density pair plots demonstrating the correlations
    between position an time delay. Simulated values are shown via
    grey lines.}
  \label{fig:hist2}
\end{figure*}

\begin{figure}[ht]
  \centering
  \begin{subfigure}[]{\includegraphics[]{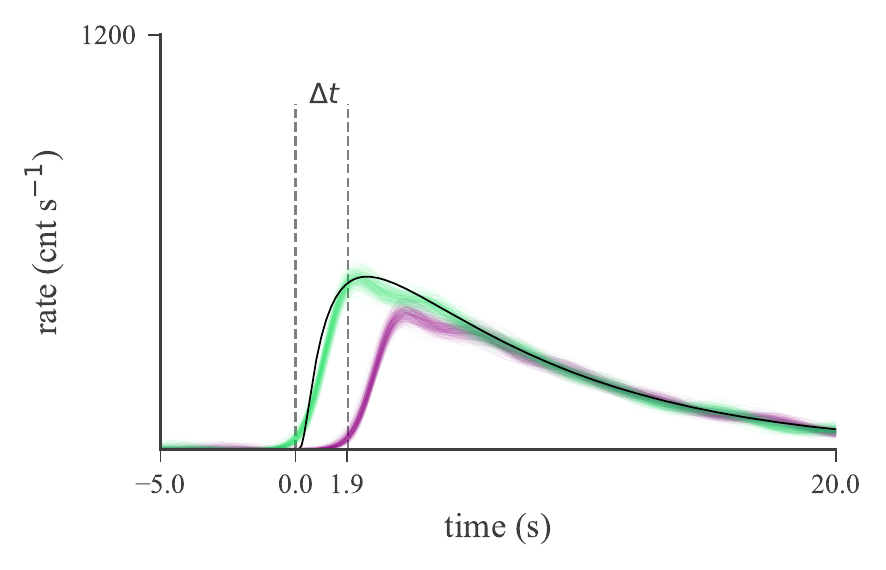}}
  \end{subfigure}
    \begin{subfigure}[]{\includegraphics[]{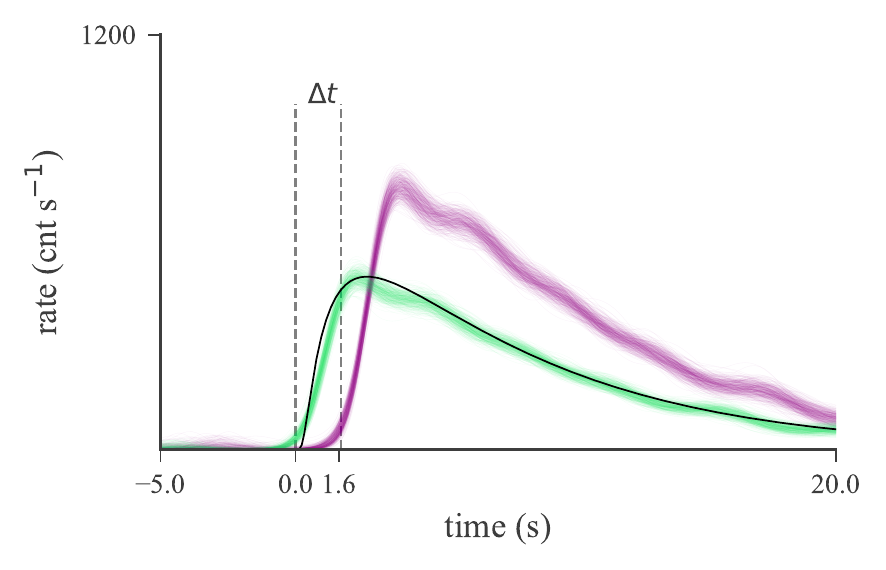}}
    \end{subfigure}
    \caption{The inferred posterior traces of the pulse models in the
      two detectors (green and purple) compared with the non-delayed
      simulated pulse model for each of the detectors (left d$_{1,2}$
      and right d$_{1,3}$).  }
  \label{fig:pure12}
\end{figure}

We verify that the light curves of the simulated data are properly
inferred via PPCs (see Figure \ref{fig:ppc2}) and can examine the
inferences of the latent pulse signal (see Figure
\ref{fig:pure12}). We again can examine the natural correlation
between position on the time delays in Figure \ref{fig:hist2}.
Similar to the two detector geometry, the inference of the latent
pulse shapes are not perfect owing to the Poisson realizations of the
data. Nevertheless, the reconstruction is good enough to infer the
time-delays between the signals.

\begin{figure}[ht]
  \centering
  \begin{subfigure}[]{\includegraphics[]{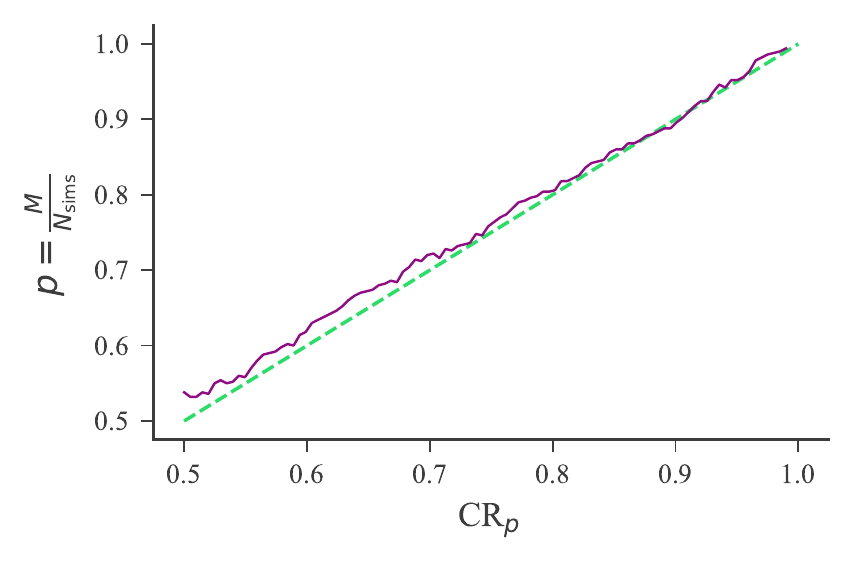}}
  \end{subfigure}
    \begin{subfigure}[]{\includegraphics[]{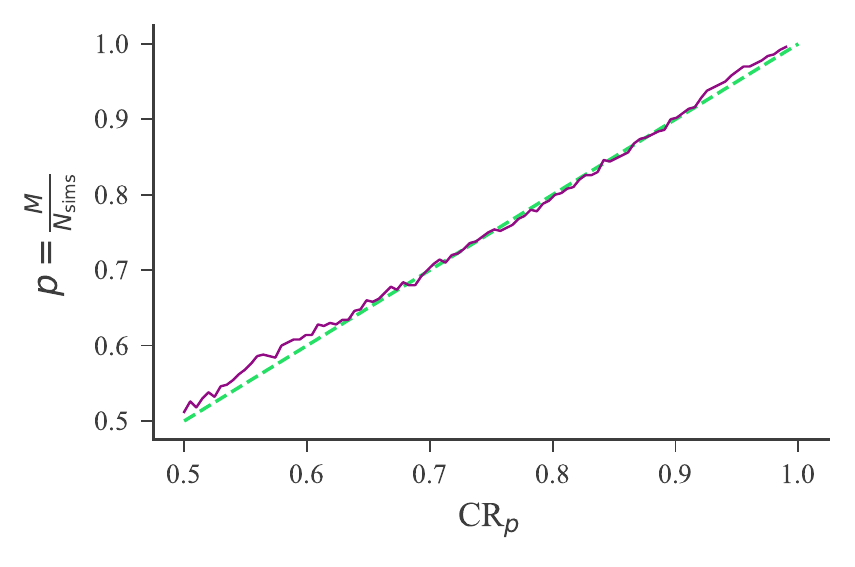}}
    \end{subfigure}
    \caption{Credible regions of $\Delta t$ versus the fraction of times
      the true, simulated value is within that credible region for
      each detector baseline. The green line represents a perfect,
      one-to-one relation. (left d$_{1,2}$
      and right d$_{1,3}$).}
  \label{fig:coverage2}
\end{figure}

\begin{figure}[ht]
  \centering
  \begin{subfigure}[]{\includegraphics[]{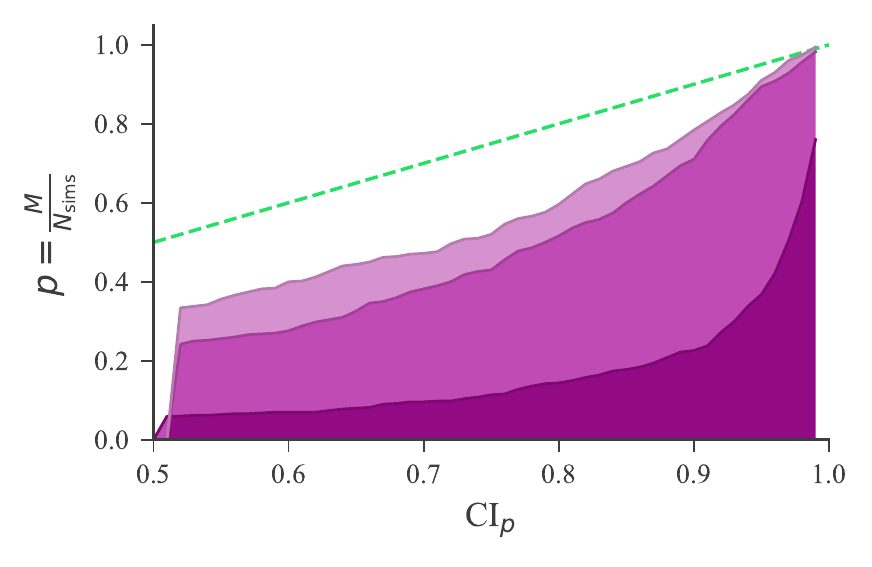}}
  \end{subfigure}
    \begin{subfigure}[]{\includegraphics[]{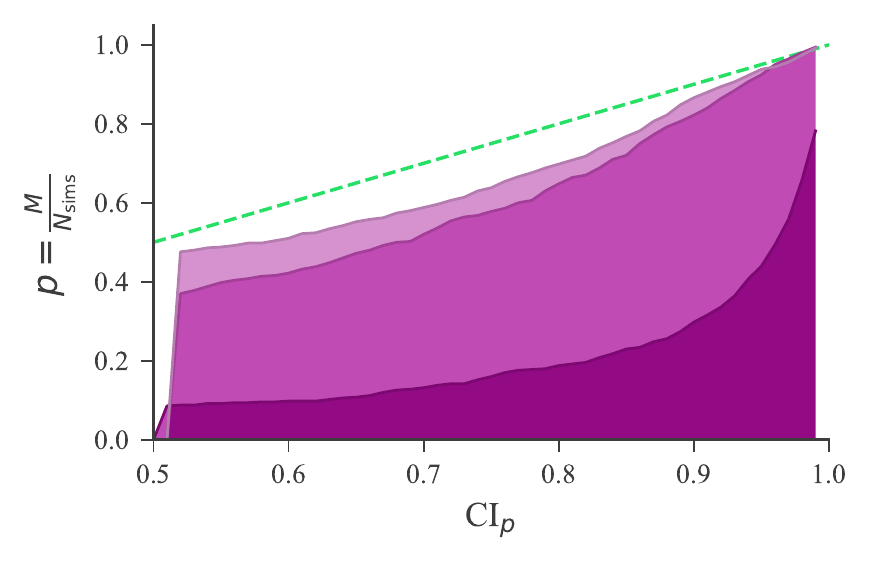}}
    \end{subfigure}
    \caption{Credible intervals of $\Delta t$ versus the fraction of times
      the true, simulated value is within that credible intervals for
      each detector baseline. The green line represents a perfect,
      one-to-one relation. Here, coarser binnings are represented by
      lighter purple colors. (left d$_{1,2}$ and right d$_{1,3}$).}
  \label{fig:coveragecc2}
\end{figure}

\begin{figure}[ht]
  \centering
  \begin{subfigure}[]{\includegraphics[]{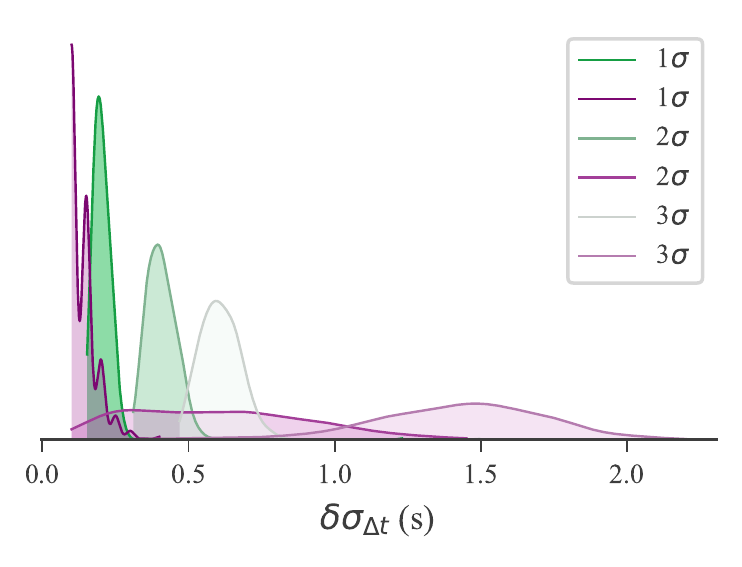}}
  \end{subfigure}
    \begin{subfigure}[]{\includegraphics[]{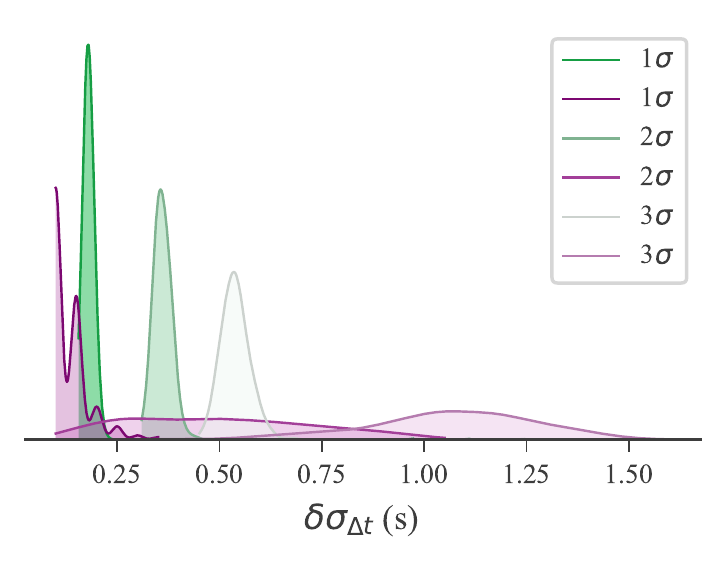}}
    \end{subfigure}
    \caption{Distributions of the relative widths of the uncertainties
    on $\Delta t$ at three different credible levels from the classical
    (purple) and new (green) algorithm fits to simulations. (left d$_{1,2}$
      and right d$_{1,3}$).}
  \label{fig:width2}
\end{figure}

The inferred uncertainties on both time delays are also robust as we
saw in the two detector case (see Figure
\ref{fig:coverage2}). However, we find that the same is again not true
from the cIPN as shown in Figure \ref{fig:coveragecc2}. Additionally,
we find the same pattern of relative widths of the of the uncertainty
regions for each of the baselines as was found in the two-detector
case (see Figure \ref{fig:width2}).

\section{Discussion}
\label{sec:discussion}
We have presented a new approach to GRB localization via signal
triangulation that demonstrates a forward modeling approach that is
statistically principled. The approach does not require erroneous
background subtraction, and is flexible enough to be improved upon
with modifications that fully encompass the complications of real
data. While the algorithm is much slower than the classical approach,
IPN localizations are already much slower than other methods due to
the latency of obtaining the data from multiple observatories, thus,
we suspect little impact to the overall localization procedure. The
computational time is; however, significant, typically requiring
$\sim 30$ minutes on a 40 core high-performance server. The runtime is
influenced by the resolution of the light curves, but only
mildly. Future improvements to the code will focus on decreasing this
runtime.  We have not addressed the issues of temporally varying
backgrounds or spectrally evolving signals when they are detected by
detectors with 1) different spectral windows and 2) high amounts of
energy dispersion. We comment on each of these issues briefly below.

\subsection{Temporally varying background}

Detectors in low-Earth orbits such as Fermi/GBM operate in the
presence of different background components that vary drastically
during their orbits due to changing Earth/sky ratio, geomagnetic
fields and radiations belts. Thus, our simple assumption of a constant
background would breakdown. Nevertheless, we can substitute this
constant assumption with a temporal model such as a polynomial or even
a secondary RFF. This will likely increase the uncertainty in the
reconstructed localization, but will be statistically robust compared
to background subtraction. We have made preliminary progress in this
direction, but defer this work to a future study.

\subsection{Spectrally evolving signals}

For detectors with the same spectral window and energy-dependent
effective area, the assumption that the time-varying signal observed
in the data is the same is valid. However, if either of these two
properties are not realized, then a more complex modeling is
required. The simplest situation is when the detectors observe the
same spectral window, but vary in their overall effective area. In
this case, the latent signal can be multiplied by the known or modeled
ratio of areas. In any other situation, a model for the spectrally
evolving signal must be parameterized such that its shape can be
fitted for and consequently, this shape must be folded through each
detectors' response. A similar approach is taken for the static case
with the BALROG algorithm \citep{Burgess:2017aa,Berlato:2019aa}. While
the approach is straight forward to implement, it will introduce
computational expense to the algorithm. Moreover, the approach would
require that the response functions of each instrument be publicly
available and well calibrated. This extra spectral information would
also enhance the ability to localize GRBs by adding an extra dimension
to match across detectors. In the same way that BALROG uses the
changing effective area of Fermi/GBM detectors to localize GRBs via
their spectral shape, the same localization capability would work in
tandem with the matching of time-differences in well separated
detectors. Thus, we will pursue this issue in future work.

\begin{figure}[ht]
  \centering
  \includegraphics[]{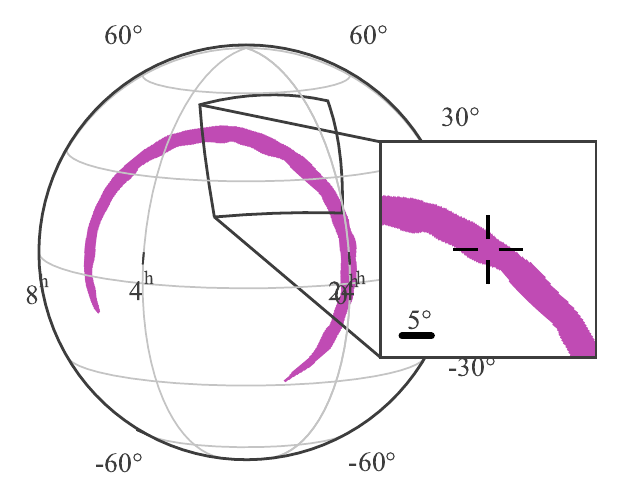}
  \caption{Sky map of the two detector geometry when the detector is
    assumed to have an effective area that is proportional to the
    angle of the GRB w.r.t. its optical axis. Only the 1$\sigma$ contour is
    shown for visual clarity.}
  \label{fig:cos}
\end{figure}

A current approximation to this issue could be to assume that at
low-energies, where most of the signal is absorbed in the detector,
the angle-dependent effective area evolves as a cosine from the
optical axis of the detector. Thus, the change in amplitude of signal
between various detectors is a function of the GRB position relative
to the optical axis. To demonstrate this we have repeated the two
detector validation but this time given each detector a cosine
dependent effective area and adopted our algorithm
appropriately. Thus, we set the amplitude of the signal arriving at
the detector such the $\tilde{A}_i = A_i \cos{\gamma}$ where
$ \gamma $ is the angle between the GRB and the optical axis of the
detector. As can be seen in Figure \ref{fig:cos}, the ring degeneracy
of the triangulation method is broken and a more precise localization
is achieved.

\subsection{Refinements to statistical model and posterior approximation}
Two theoretical issues with the proposed statistical model are worth
noting, although they do not appear to limit the performance of the
{\tt nazgul} algorithm in any of the case studies examined in this simulation
study.  The first is that the credible intervals in Bayesian inference
are not universally guaranteed to achieve Frequentist style coverages.
That is, a 90\% posterior credible interval is not guaranteed to offer
at least a 90\% coverage of the true parameter value in the long-run
over mock datasets as would an `exact' Frequentist interval.  The
risks to the reliability of astronomical inference procedures due to
supposing \textit{a priori} that Bayesian intervals offer such
coverage guarantees have been noted by
\citet{Genovese:2004aa}, but remain little appreciated in
the community.  The model type studied here in which the target of
interest is in effect a functional of a non-parametric stochastic
process prior is also similar to one in which pathological asymptotic
results have been demonstrated in theoretical studies
\citep{Castillo:2012aa}.  For this reason we emphasise the
importance of validation by simulated data in general, and in
particular as further model complexity is added to increase the range
of GRB observing scenarios covered by the {\tt nazgul} algorithm.

The other theoretical issue of potential concern is with respect to
behaviour of Gaussian processes built on the squared exponential
kernel.  In particular, it is known that this kernel favours an
aggressive smoothing which can lead to poor Frequentist style coverage
of the latent function being approximated \citep{Hadji:2019aa}. In
this case we in fact use a non-stationary version of the squared
exponential kernel which allows for a non-universal degree of
smoothness, which might reduce exposure to this problem. On the other
hand, it is evident that our chosen kernel and approximation via a
small number of random Fourier features does lead to some degree of
over-smoothing at the sharpest transition points, such as prior to
initial burst up-tick in Figure \ref{fig:simonefit}.  Again, this does
not seem to limit the accuracy of time delay recoveries, and there are
possible advantages of a smooth kernel that might be working in our
favour. For instance, a smoother kernel is less likely to attribute a
cosmological origin to high background noise spikes, and can be better
approximated with fewer random Fourier features than a less smooth
kernel.

The current approach to posterior approximation is via MCMC simulation
under a Hybrid Monte Carlo scheme (implemented in \texttt{Stan}).
While this method offers a convergent approximation of the exact
Bayesian posterior, it is in this case notably expensive in terms of
computational resources and runtime.  In similar problems it has been
shown that first order approximations to the Bayesian posterior can be
both adequate for uncertainty quantification and much more
computationally efficient.  For instance, in the setting of estimating
time delays in gravitationally lensed quasar light curves,
\citet{Tak:2017aa} identified substantial computational gains
from a profile-likelihood based approach.  In our experiments with the
\texttt{nazgul} algorithm we have been able to confirm an order of
magnitude speed up in posterior sampling when the bandwidth parameter
is held fixed, suggesting that a promising avenue of investigation may
be to search for a heuristic bandwidth selection strategy to set this
parameter to a reasonable value at the start of model fitting.  An
alternative may be to adapt the mesh-based approximation to Gaussian
process inference offered by the \texttt{INLA} package
\citep{Lindgren:2010aa}, which can also accommodate
non-stationarity and offers speed ups equivalent to profile-likelihood
based approaches.  In either case it will be important to establish
that any loss of accuracy in the uncertainty estimation stays within
an acceptable tolerance to avoid losing the demonstrated coverage
properties of the full posterior sampling scheme.

\subsection{Methods for model misspecification}
Finally, it is worth noting here that there are a number of methods,
both established and emerging, for reducing the impact of model
misspecification on parameter inference in either the Bayesian and
Frequentist setting which may be worthwhile exploring in this context.
Model misspecification occurs when the process from which the observed
data are generated is not within the parametric family assumed by the
modeler.  In such cases the behaviour of likelihood-based approaches
is typically to converge towards the set of input parameters that
minimise the Kullback-Liebler divergence with respect to the true
sampling distribution.  Depending on the nature of the
misspecification this parameter set may be the same as, or close to,
that which would be targeted if the assumed model were correct, but
the behaviour of the associated credible or confidence intervals
associated with misspecified estimates tends to remain asymptotically
incorrect.  Methods such as Bayesian bagging \citep{Huggins:2019aa},
the weighted likelihood boostrap \citep{Spokoiny:2015aa}, and
the Bayesian bootstrap \citep{Lyddon:2019aa} can correct the
tendency of these intervals to offer coverages lower than their
nominal levels.  While these three methods have been developed and
studied almost exclusively under the iid (independent, identically
distributed) data setting for simple parametric models, there is
ongoing interest in extension to more complicated settings for which
the GRB localization problem may prove a compelling case study.

\section{Conclusion}
\label{sec:conclusion}

The use of GRB IPN triangulations has been a successful enterprise,
benefiting both the GRB and gravitational wave communities. However,
we have shown that statistical uncertainties of the cIPN lack
robustness.  With our {\tt nazgul} algorithm and the associated
publicly available
code\footnote{\faGithub~\href{https://github.com/grburgess/nazgul}{https://github.com/grburgess/nazgul}},
we intend to work with the communities to increase the fidelity of the
IPN with the aim of providing robust, statistically sound
localizations in the on-going multi-messenger era. We find that the
method provides better accuracy and precision. We foresee that issues
will arise when dealing with real data, however, these issues exist
for the cIPN as well. Given the success of the cIPN to locate many
GRBs, we believe that these issues will be mitigated. It was brought
to our attention (K. Hurley, \textit{private communication}), that
attempts to obtain time-delays via the fitting of functional forms to
the data of the Vela satellites \citep{Strong:1974aa} though this work
was never published.

There are also technical benefits to our approach due to the fact that
Bayesian inference produces probabilities directly. First, our
localization marginals can be directly convolved with other posteriors
such as those of BAYESTAR \citep{Singer:2016aa}, BALROG
\citep{Burgess:2017aa}, or any tool that also produces a localization
posterior. Along these lines, we can also distribute the localizations
in either HEALPix \citep{Gorski:2005aa} or the the newer multi-order
coverage maps \citep[MOC,][]{Fernique:2015aa} allowing rapid and
direct access to the underlying location information without the need
for pseudo localization combination techniques.

We further note that the methods developed here extend beyond GRB
triangulation and can be used for other time-difference based studies
which rely on signal variability with no a priori model such as the
micro-lensing in quasars for the determination of dark matter content
\citep{Suyu:2017aa} or for analysis of reverberation mapping of active
galactic nuclei \citep{Wilkins:2019aa}. Even in cases of an existing
a priori model, our methods might be advantageous if observing times
are too short to beat the (timing) noise, such as X-ray millisecond
pulsar navigation \citep{Becker:2018aa}. Finally, these methods could
be used in the search for gravitationally lensed GRBs
\citep{Grossman:1994aa, Hurley:2019aa} as they do not require the
search for any a priori defined pulse shape or the use of
cross-correlations.
\section*{Software}
\label{sec:software}

All software used in this publication including the framework and
analysis scripts are publicly available. The work was made possible
via {\tt Astropy} \citep{Price-Whelan:2018aa}, {\tt matplotlib}
\citep{Hunter:2007aa}, {\tt numpy} \citep{Harris:2020aa}, {\tt SciPy}
\citep{Virtanen:2020aa}, {\tt numba} \citep{Lam:2015aa}, {\tt Stan}
\citep{Carpenter:2017aa}, {\tt arviz} \citep{Kumar:2019aa}, and {\tt
  ligo.skymap} \citep{Singer:2016aa}.

\begin{acknowledgements}
  The authors are thankful for conversations with Kevin Hurley,
  Valentin Pal'shin, Alex Nitz, Aaron Tohuvavohu and Thomas Siegert,
  in helping make the manuscript clearer. We thank the Max-Planck
  Computing and Data Facility for the use of the Cobra HPC
  system. Additionally, we thank Leo Singer for pointing us in the
  right direction to make sky maps. JMB acknowledges support from the
  Alexander von Humboldt Foundation.

\end{acknowledgements}

\bibliographystyle{aa}
\bibliography{bib.bib}
\appendix

\section{Simulations}
\label{sec:sims}
To demonstrate the method, we designed a simulation package that
allows us to create count light curves from any configuration of GRB
sky position, and spacecraft orbit. To establish the geometry of the
simulation a configuration file in YAML \citep{} format is used. The
user specifies the sky location of the GRB and its distance from the
Earth. Then, for each GRB detector, the location is specified as the
sky location and altitude above the Earth. The detector's optical axis
or pointing direction is specified as a sky coordinate along with the
detector's effective area in cm$^2$. For simplicity, the effective
area decreases as the cosine of the angle from the pointing
direction's normal vector. An example is given below:

\begin{Verbatim}[commandchars=\\\{\}]
\PYG{c+c1}{\PYGZsh{} Specify the GRB parameters}
\PYG{n+nt}{grb}\PYG{p}{:}

   \PYG{c+c1}{\PYGZsh{} Location and distance (degrees and Mpc)}
   \PYG{n+nt}{ra}\PYG{p}{:} \PYG{l+lScalar+lScalarPlain}{20.}
   \PYG{n+nt}{dec}\PYG{p}{:} \PYG{l+lScalar+lScalarPlain}{40.}
   \PYG{n+nt}{distance}\PYG{p}{:} \PYG{l+lScalar+lScalarPlain}{500.}

   \PYG{c+c1}{\PYGZsh{} lightcurve}
   \PYG{c+c1}{\PYGZsh{} if arrays are used then}
   \PYG{c+c1}{\PYGZsh{} multiple pulses are created}
   \PYG{n+nt}{K}\PYG{p}{:} \PYG{p+pIndicator}{[}\PYG{n+nv}{400.}\PYG{p+pIndicator}{,}\PYG{n+nv}{400} \PYG{p+pIndicator}{]} \PYG{c+c1}{\PYGZsh{} intensity}
   \PYG{n+nt}{t\PYGZus{}rise}\PYG{p}{:} \PYG{p+pIndicator}{[}\PYG{n+nv}{.5}\PYG{p+pIndicator}{,} \PYG{n+nv}{.5}\PYG{p+pIndicator}{,]} \PYG{c+c1}{\PYGZsh{} rise time}
   \PYG{n+nt}{t\PYGZus{}decay}\PYG{p}{:} \PYG{p+pIndicator}{[}\PYG{n+nv}{4}\PYG{p+pIndicator}{,} \PYG{n+nv}{3}\PYG{p+pIndicator}{,]} \PYG{c+c1}{\PYGZsh{} decay time}
   \PYG{n+nt}{t\PYGZus{}start}\PYG{p}{:} \PYG{p+pIndicator}{[}\PYG{n+nv}{0.}\PYG{p+pIndicator}{,} \PYG{n+nv}{4.}\PYG{p+pIndicator}{]} \PYG{c+c1}{\PYGZsh{} start time}

\PYG{c+c1}{\PYGZsh{} Specify the detectors}
\PYG{c+c1}{\PYGZsh{} each entry is treated as}
\PYG{c+c1}{\PYGZsh{} the name of the detector}
\PYG{n+nt}{detectors}\PYG{p}{:}

    \PYG{n+nt}{det1}\PYG{p}{:}

        \PYG{c+c1}{\PYGZsh{} 3D position is separated into}
        \PYG{c+c1}{\PYGZsh{} sky location (GCRS) and altitude}
        \PYG{n+nt}{ra}\PYG{p}{:} \PYG{l+lScalar+lScalarPlain}{40.}
        \PYG{n+nt}{dec}\PYG{p}{:} \PYG{l+lScalar+lScalarPlain}{5.}
        \PYG{n+nt}{altitude}\PYG{p}{:} \PYG{l+lScalar+lScalarPlain}{1500000.} \PYG{c+c1}{\PYGZsh{} km}
        \PYG{n+nt}{time}\PYG{p}{:} \PYG{l+s}{\PYGZsq{}2010\PYGZhy{}01\PYGZhy{}01T00:00:00\PYGZsq{}}

        \PYG{c+c1}{\PYGZsh{} optical axis of the detector}

        \PYG{n+nt}{pointing}\PYG{p}{:}
            \PYG{n+nt}{ra}\PYG{p}{:} \PYG{l+lScalar+lScalarPlain}{20.}
            \PYG{n+nt}{dec}\PYG{p}{:} \PYG{l+lScalar+lScalarPlain}{40.}

        \PYG{n+nt}{effective\PYGZus{}area}\PYG{p}{:} \PYG{l+lScalar+lScalarPlain}{1.}

    \PYG{n+nt}{det2}\PYG{p}{:}

       \PYG{n+nt}{ra}\PYG{p}{:} \PYG{l+lScalar+lScalarPlain}{60.}
       \PYG{n+nt}{dec}\PYG{p}{:} \PYG{l+lScalar+lScalarPlain}{10.}
       \PYG{n+nt}{altitude}\PYG{p}{:} \PYG{l+lScalar+lScalarPlain}{500}
       \PYG{n+nt}{time}\PYG{p}{:} \PYG{l+s}{\PYGZsq{}2010\PYGZhy{}01\PYGZhy{}01T00:00:00\PYGZsq{}}

       \PYG{n+nt}{pointing}\PYG{p}{:}

           \PYG{n+nt}{ra}\PYG{p}{:} \PYG{l+lScalar+lScalarPlain}{20.}
           \PYG{n+nt}{dec}\PYG{p}{:} \PYG{l+lScalar+lScalarPlain}{40.}

       \PYG{+Error}{ }\PYG{n+nt}{effective\PYGZus{}area}\PYG{p}{:} \PYG{l+lScalar+lScalarPlain}{.5}
\end{Verbatim}

The remaining information is the shape of the light curve. The shape
we use for the simulation is that of \citet{Norris:1996aa} which
specifies a simple pulse that rises and decays in time.
\begin{equation}
  \label{eq:2}
  s(t; K, \tau_{\mathrm{r}}, \tau_{\mathrm{d}}) = K e^{2 \sqrt{\tau_{\mathrm{r}}/ \tau_{\mathrm{d}}}} e^{-\tau_{\mathrm{r}}/t - t/\tau_{\mathrm{d}}}
\end{equation}

\noindent
Here, $K$ is the amplitude of the pulse at the peak,
$\tau_{\mathrm{r}}$ is the rise time constant, and $\tau_{\mathrm{d}}$ is
the decay time constant. Any synthetic GRB within the framework can be
composed of multiple Norris-like pulses.

Once the user specifies the parameters, the detectors are ordered by
the arrival time of the GRB photons and a time-system is established
between all the detectors such that the T$_0$ or reference time is the
arrival of the first photon at the detector closest to the GRB. Once
this time system is established, individual photon events are sampled
according to an inhomogeneous-Poisson distribution following the
intrinsic pulse shape specified and weighted by each detectors'
effective area. The photons arrival times are sampled via an inverse
cumulative distribution function (CDF) rejection sampling scheme
\citep{Rubinstein:2016aa}.

\begin{figure}[ht]
  \centering
  \includegraphics[]{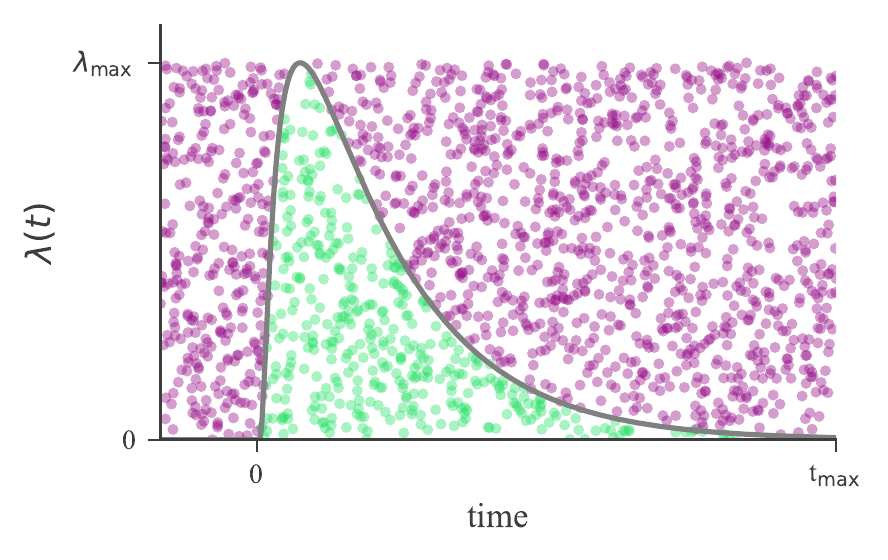}
  \caption{A demonstration of the rejection sampling from a
    Norris-like pulse shape where accepted samples are shown in green
    and rejected samples are shown in purple}
  \label{fig:rejection}
\end{figure}

Assume we want to know how long one must wait ($T$) until a
new photon arrives when our counts are Poisson distributed with
constant rate $\lambda$.

\begin{equation}
  \label{eq:3}
  \pi(T>t)=\pi(f(t)=0)=\frac{(\lambda t)^{0} e^{-\lambda t}}{0 !}=e^{-\lambda t}
\end{equation}
\noindent
which tells us that the waiting times are exponentially
distributed. Thus, we need to inverse CDF sample the exponential
distribution to sample arrival times of photons. The CDF of the
exponential distribution is
\begin{equation}
  \label{eq:4}
  F(t)=1-e^{-\lambda t}
\end{equation}
\noindent
thus we need to solve
\begin{equation}
  \label{eq:5}
  F\left(F^{-1}(u)\right)=u
\end{equation}
\noindent
where $u \sim U(0,1)$. This yields

\begin{equation}
  \label{eq:6}
  t=F^{-1}(u)=-\frac{1}{\lambda} \log (1-u)
\end{equation}
\noindent
which is invariant to translations of $u$ and thus we arrive at the
equation to sample for the time between events.

\begin{equation}
  \label{eq:7}
  t=F^{-1}(u)=-\frac{1}{\lambda} \log (u) \text{.}
\end{equation}
\noindent
This allows us to sample a constant rate as is done for our
background. We start sampling at the start time of the light curve,
and continue until a specified $t_{\mathrm{max}}$.  As the rate for
the signal evolves with time, we implement a further rejection
sampling step that thins the arrival times according to the evolving
$\lambda(t) = s(t; K, \tau_{\mathrm{r}}, \tau_{\mathrm{d}})$. This is done by
first sampling a waiting time $t$ and computing the $\lambda(t)$. A draw
from $p \sim U(0,1)$ is made and the sample is accepted if
$ p\le \lambda(t)$ as shown in Figure \ref{fig:rejection}.

Thus, the simulated events observed in each detector can be
arbitrarily binned into any required cadence for the purpose of our
analysis. Code and examples are available
\faGithub\href{https://github.com/grburgess/pyipn}{~here}.

\section{Fitting algorithm} \label{sec:fitting}
The high-dimensionality of this model requires a sampling algorithm
capable of constructing the posterior accurately. For this, we choose
the enhanced NUTS algorithm \citep{Betancourt:2017ab} implementation
in Stan \citep{Carpenter:2017aa}. For every chain, we use 1000 warmup
and 1000 sample iterations.

For the parameterization of the RFFs, we note that sampling for
$\omega_{1,2}$ provided more robust estimates than choosing them
randomly. However, placing a prior of
$\omega_{1,2} \sim\mathcal{N}(0, b_{1,2})$ creates a one-way normal
\citep{Betancourt:2013aa} resulting in a funnel geometry. Thus, we choose a
non-centered parameterization such that
$\tilde{\omega}_{1,2} \sim \mathcal{N}(0,1) $ and
$\omega_{1,2} = \tilde{\omega}_{1,2} \cdot b$. We also found that a more efficient
parameterization which resulted in no divergences required sampling
the scale-length ($l$) or inverse of $b$ normalized to the
boundaries of the time-series. This removed divergent transitions near
the boundaries of the $b$ marginal distributions. 

\begin{itemize}
\item The integer count data, space craft position vectors in GCRS,
  and number of random features are loaded into Stan
\item The GRB location prior is parameterized as a unit Cartesian
  3-vector i.e., uniform on the sky.
\item $\vec{\tilde{\omega}}_{1,2}$ are given unit priors
\item $l$ are sampled from weakly-informative priors bounded
  but the length of the time-series and $b = \frac{1}{l}$.
\item $\vec{\beta}_{1,2}$ are give unit normal priors
\item all scale parameters are given log normal priors
\item a given detector is set as the reference detector from which all
  time-differences are computed. For each sample, the time delays are
  computed given the the sampled GRB sky position.
\item
  $S(t; b_{1,2},\vec{\omega}_1,\vec{\omega}_2, \vec{\beta}_1, \vec{\beta}_2 )$ is computed for
  the reference detector and the time differences for all other
  detectors are used to compute the delayed signal.
\item the background level is estimated 
  
\item each detectors' signal and background are summed and compared to
  the observed counts via a Poisson likelihood
  
\end{itemize}

In order to increase the speed of the algorithm, we have spilt the
computation of the likelihood across several threads via Stan's {\tt
  reduce\_sum} functionality. We have observed that the problem scales
quasi-linearly with the number of threads. All fits checked for
convergence, high effective sample size, and required to be divergence
free. We note that sporadic divergences appear even after
re-parameterization but can be removed by slightly decreasing the
adaptive step-size. Code and examples are available
\faGithub\href{https://github.com/grburgess/nazgul}{~here}.

\section{Cross-correlation algorithm for cIPN}
\label{sec:ccr}
We briefly describe our implementation of the cross-correlation
algorithm to mimic the approach taken in the cIPN \citep[for explicit
details, see e.g.][]{Palshin:2013aa,Hurley:2013aa}. Assuming we have
two light curves binned to resolution $\delta$ such that the total counts
in each light curve time are $c_{1,2}(t_i)$ = $c_{1,2; i}$. With a
known background one can subtract background counts from the total
light curve. We denote these background subtracted counts as
$\tilde{c}_{1,2}$. Now, a pseudo-statistic can be constructed such
that
\begin{equation}
  \label{eq:8}
  R^{2}(\Delta t \equiv j \delta)=\sum_{i=i_{\text {start }}}^{i=i_{\text {start }}+N} \frac{\left(\tilde{c}_{2, i}-s \tilde{c}_{1, i+j}\right)^{2}}{\left(\sigma_{2, i}^{2}+s^{2} \sigma_{1, i+j}^{2}\right)}
\end{equation}
\noindent
where $j$ indicates multiples of the base resolution. Here, $s$ is the
ratio of total counts in each light curve which is a heuristic way to
account for the effective area varying across detectors\footnote{Note
  that we fit for the changing effective area directly.}. Finally,
$\sigma^2_{1,2}= c_{1,2}$ are the ubiquitously used estimators for the
Poisson variance. Confidence intervals are then constructed via the relation
\begin{equation}
  \label{eq:9}
  R_{r, 3 \sigma}^{2}=\chi_{r, 3 \sigma}^{2}(N)+R_{r, \min }^{2}-1 \text{.}
\end{equation}
\noindent
Here, $N-1$ indicates the number of degrees of freedom which are the
number of time intervals. We can immediately identify points where
this formalism can breakdown. At fine temporal resolutions, the
estimator for the Poisson variance deviates from its central limit
theorem approximation. Moreover, if there are zero counts in the
intervals, the equation diverges. Secondly, the approximation of
$\chi^2$ rarely holds and even more so for pseudo-statistics and the
number of degrees of freedom is very dependent on the structure of the
problem \citep{Andrae:2010aa}.  Lastly, it is well-known that the
subtraction of two Poisson rates results in Skellam rather than
Poisson distributed data. Depending on how background is subtracted,
it may not even be the case that two Poisson rates are subtracted. For
example, if the background is estimated via a model fit then the
subtraction of the inferred counts must propagate the uncertainty of
these inferences which will lead to a more complex form of statistic
\citep{Vianello:2018aa}. The combination of approximations and
asymptotics required for this approach understandably lead to
uncertainties that do not provide adequate coverage.

\section{Significance}
\label{sec:sig}
\begin{figure}[h]
  \centering
  \includegraphics[]{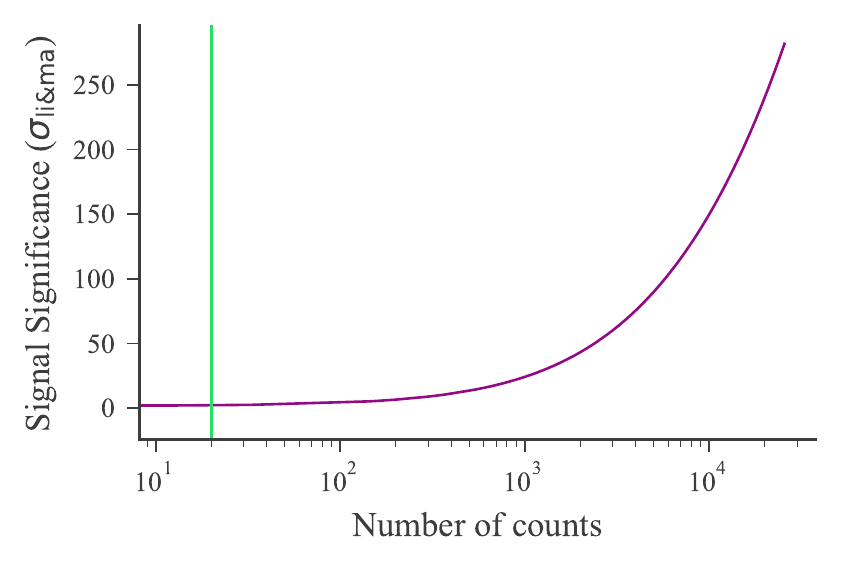}
  \caption{Significance as a function of background subtracted source
    counts.}
  \label{fig:sigcounts}
\end{figure}

It is well known \citep{Li:1983aa,Vianello:2018aa} that using simple
ratios of total and background counts rather than likelihood ratios
leads to biased estimates of significance in counting
experiments. This can also lead to overconfidence in when the
asymptotics of the central limit theorem are reached. Therefore, in
this work we employ the measure of significance derived in
\citet{Li:1983aa} where we measure the counts in a temporally
off-source region and compare it to the total counts at the peak of
the pulse and adjust for the exposure differences. We can examine the
significance as a function of background subtracted counts in Figure
\ref{fig:sigcounts} where we have subtracted the total observed counts
by the true background rate multiplied by the exposure.

\end{document}